\documentclass[10pt,twocolumn]{article} 
\usepackage{simpleConference}
\usepackage{times}
\usepackage{graphicx}
\usepackage{amssymb}
\usepackage[T1]{fontenc} 
\usepackage{authblk}
\usepackage[ruled,vlined]{algorithm2e}
\usepackage[dvipsnames]{xcolor}

\usepackage{graphicx}
\usepackage{amsmath}
\usepackage{caption}
\usepackage{multirow} 
\usepackage{multicol}
\usepackage{xcolor} 

\usepackage{tabularx}
\usepackage{subcaption}
\usepackage{array}

\let\titleold\title
\renewcommand{\title}[1]{\titleold{#1}\newcommand{\thetitle}{#1}}
\def\maketitlesupplementary
   {
   \newpage
       \twocolumn[
        \centering
        \Large
        \textbf{\thetitle}\\
        \vspace{0.5em}Supplementary Material \\
        \vspace{1.0em}
       ] 
   }

\definecolor{iccvblue}{rgb}{0.21,0.49,0.74}
\usepackage[pagebackref,breaklinks,colorlinks,allcolors=iccvblue]{hyperref}
\usepackage[top=0.5in, bottom=1in, left=0.5in, right=0.5in]{geometry}

\begin{document}

\title{CADD: Context aware disease deviations via restoration of brain images using normative conditional diffusion models}

\author[1,3]{Ana Lawry Aguila}
\author[2,3]{Ayodeji Ijishakin}
\author[1,3,4]{Juan Eugenio Iglesias}
\author[5]{Tomomi Takenaga}
\author[6]{Yukihiro Nomura}
\author[7]{Takeharu Yoshikawa}
\author[5]{Osamu Abe}
\author[5]{Shouhei Hanaoka}

\affil[1]{Athinoula A. Martinos Center for Biomedical Imaging, Massachusetts General Hospital and Harvard Medical School, Boston, USA}
\affil[2]{Mecha Health, San Francisco, USA}
\affil[3]{Hawkes Institute, University College London, London, UK}
\affil[4]{Computer Science \& Artificial Intelligence Lab, Massachusetts Institute of Technology, Boston, USA}
\affil[5]{Department of Radiology, the University of Tokyo, Tokyo, Japan}
\affil[6]{Medical Engineering, Chiba University, Chiba, Japan}
\affil[7]{Department of Computational Diagnostic Radiology and Preventive Medicine, the University of Tokyo Hospital, Tokyo, Japan}

\affil[ ]{\vspace{2em}\texttt{acaguila@mgh.harvard.edu}}

\maketitle
\begin{abstract}
Applying machine learning to real-world medical data, e.g. from hospital archives, has the potential to revolutionize disease detection in brain images. However, detecting pathology in such heterogeneous cohorts is a difficult challenge. Normative modeling, a form of unsupervised anomaly detection, offers a promising approach to studying such cohorts where the ``normal'' behavior is modeled and can be used at subject level to detect deviations relating to disease pathology. Diffusion models have emerged as powerful tools for anomaly detection due to their ability to capture complex data distributions and generate high-quality images. Their performance relies on image restoration; differences between the original and restored images highlight potential abnormalities. However, unlike normative models, these diffusion model approaches do not incorporate clinical information which provides important context to guide the disease detection process. Furthermore, standard approaches often poorly restore healthy regions, resulting in poor reconstructions and suboptimal detection performance. We present CADD, the first conditional diffusion model for normative modeling in 3D images. To guide the healthy restoration process, we propose a novel inference inpainting strategy which balances anomaly removal with retention of subject-specific features. Evaluated on three challenging datasets, including clinical scans, which may have lower contrast, thicker slices, and motion artifacts, CADD achieves state-of-the-art performance in detecting neurological abnormalities in heterogeneous cohorts.

\end{abstract}    

\section{Introduction}
\begin{figure}
    \centering
    \setlength{\abovecaptionskip}{3pt} 
    \setlength{\belowcaptionskip}{0pt} 
    \includegraphics[trim={0 0 0 0}, clip, width=0.9\columnwidth]{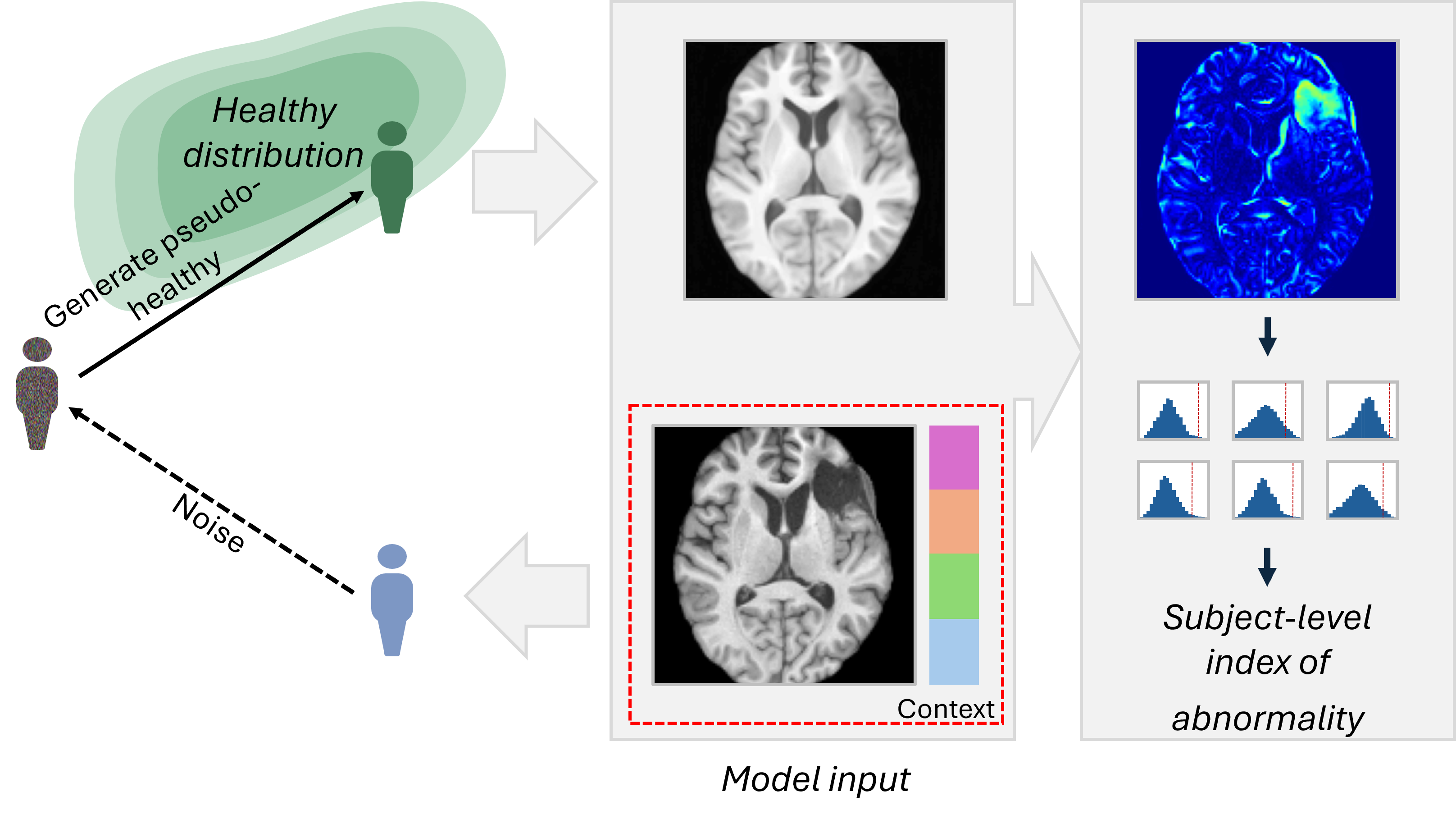}
    \caption{Using a diffusion model-based normative framework for disease detection.}
    \label{fig:CADD_overview}
\end{figure}

Machine learning has the potential to transform disease detection in clinical data. However, disease heterogeneity and data availability present significant challenges in the study of neurological diseases. Large, real-world datasets often contain disease labels which are poorly defined, if available at all, and encompass a variety of disease types. Normative modeling is a type of Out-of-Distribution (OoD) or anomaly detection for describing the ``normal'' behavior of a healthy population which can be used at subject level to detect deviations relating to a disease. Unlike standard anomaly detection, normative models incorporate confounding covariates (e.g., age, sex) to avoid obscuring or inflating pathological effects. However, traditional normative approaches \cite{Marquand:2016} do not take into account the interactions between features and are computationally unfeasible for large brain imaging datasets, which often contain millions of voxels. Recently, to model complex non-linear interactions between features, deep-learning approaches using autoencoder models have been proposed \cite{Pinaya:2021,LawryAguila:2022}. For measuring deviations in the feature space, these approaches consider the reconstruction error between the original and reconstructed data. 

Such reconstruction-based methods have become popular in the wider field of anomaly detection \cite{Denouden2018,Baur2020,Somepalli2021,Gong2019,zong2018}. Commonly, these methods are trained in an unsupervised manner requiring only in-distribution data for training, with anomalies being detected from inaccurate reconstructions of anomalous samples. However, a number of works have highlighted issues with reconstruction-based methods \cite{Bercea2023,cai2024,Denouden2018}. In particular, using a sufficiently constrained latent space for anomaly detection, comes at the cost of low quality reconstructions which compromises the utility of such reconstructions for downstream tasks. 

Diffusion models \cite{Ho2020,sohldickstein2015} have achieved state-of-the-art results in generative modeling and have recently outperformed other generative models in anomaly detection in brain imaging \cite{Pinaya2022,Graham2023b, Wolleb2022}.
By modeling data distributions through fixed Gaussian noising and learnable denoising steps, diffusion models capture more expressive representations of complex data compared to previous generative methods. Due to the computational challenges of using 3D brain images, many approaches use a Latent Diffusion Model (LDM) \cite{Rombach2021} where a first stage autoencoder mode (with a large bottleneck and thus a sufficiently expressive latent space) is used to reduce the dimensionality of the input data to a latent space on which the diffusion model is trained. As far as we are aware, there has been no application of LDMs for anomaly detection in 3D MRI images of common neurological diseases with prior work focusing on detecting artificial or large brain lesions \cite{Graham2023b}. In anomaly detection, diffusion models trained on healthy data are used to transform pathological tissue into healthy tissue by adding and then removing noise. Typically, reconstruction begins from a partially noised image to help preserve information from the original image. However, the balance between successfully removing anomalous regions whilst retaining individual level characteristics poses a challenge for diffusion model approaches \cite{graham2023c,Bercea2023b,Bercea2024c}. Here, we introduce an inference inpainting scheme which uses an element of the diffusion model training objective to identify anomalous regions during the denoising process and generate realistic, pseudo-healthy reconstructions which preserve healthy regions. This allows for the application of standard brain image segmentation or other processing algorithms which would often fail in the presence of pathology \cite{durrer2024,Kofler2024}. 

We introduce the first conditional diffusion model-based normative framework for disease detection in 3D brain images. By incorporating confounding covariates through conditioning, our approach enables reconstructions and anomaly scores to be adjusted for clinical context for the first time. We validate our model on three highly challenging brain imaging datasets which have weak disease signals, confounding factors, or have data taken directly from the clinic. These clinical scans may be of lower image contrast, thicker slices and may include motion artifacts which present additional challenges for disease detection. We make the following contributions: \textit{(i)}~We present CADD; a transformer-based normative conditional diffusion model for Context Aware Disease Detection in 3D brain images. \textit{(ii)}~We introduce an inpainting scheme, with an interative thresholding approach, at inference time to preserve healthy tissue whilst effectively removing pathological effects. \textit{(iii)}~We present the first, as far as we are aware, application of diffusion model anomaly detection to 3D brain T1-weighted MRI images from a clinical dataset, highlighting the potential of these models to be applied in clinic.


\begin{figure*}[ht]
    \centering
    \setlength{\abovecaptionskip}{3pt} 
    \setlength{\belowcaptionskip}{0pt} 
    \makebox[0.95\linewidth]{ 
        \centering
        \begin{subfigure}{0.3\textwidth}
            \centering
            \includegraphics[width=\linewidth]{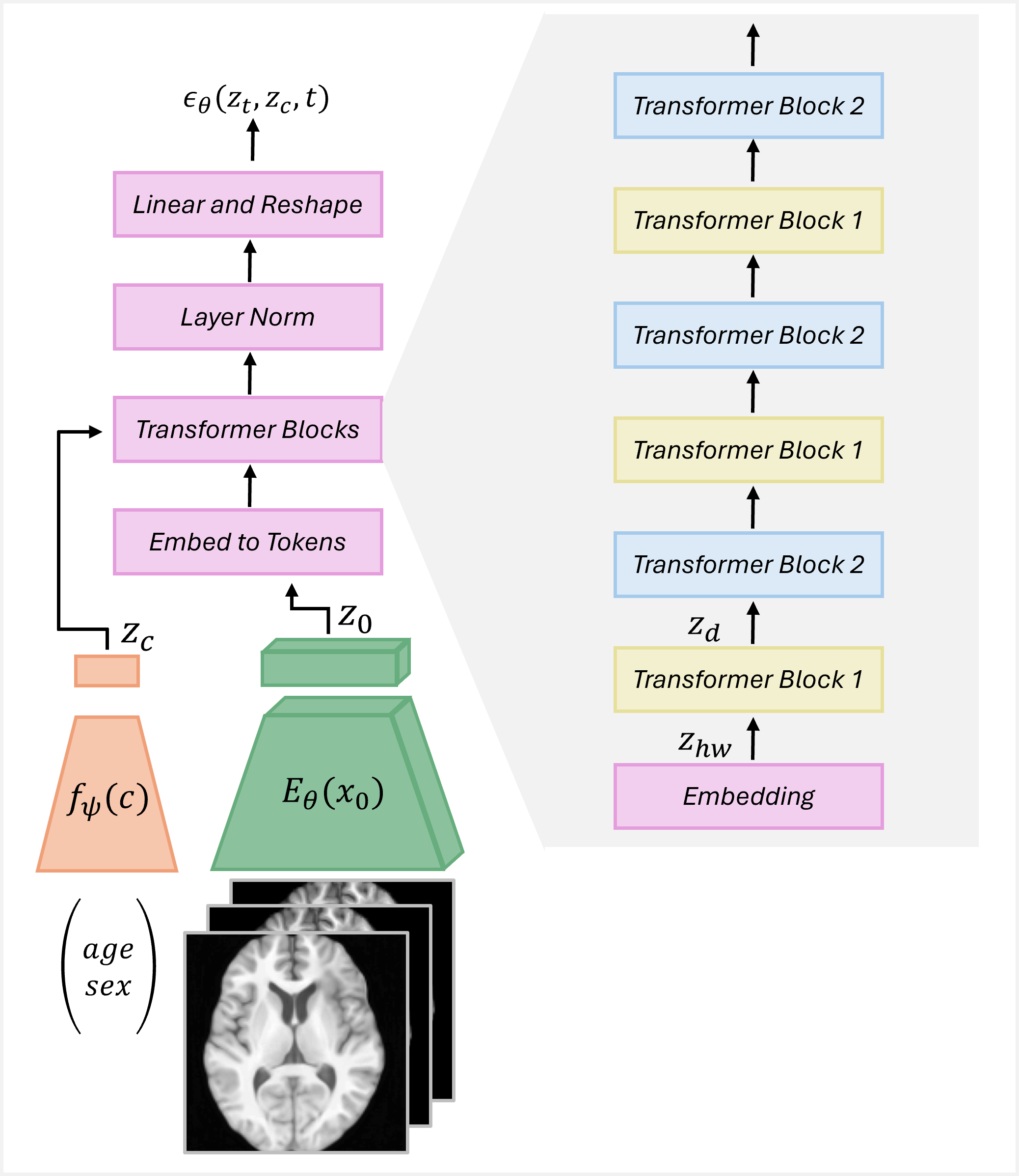}
            \caption{}
            \label{fig:subfig2}
        \end{subfigure}
        \hfill
        \begin{subfigure}{0.3\textwidth}
            \centering
            \includegraphics[width=\linewidth]{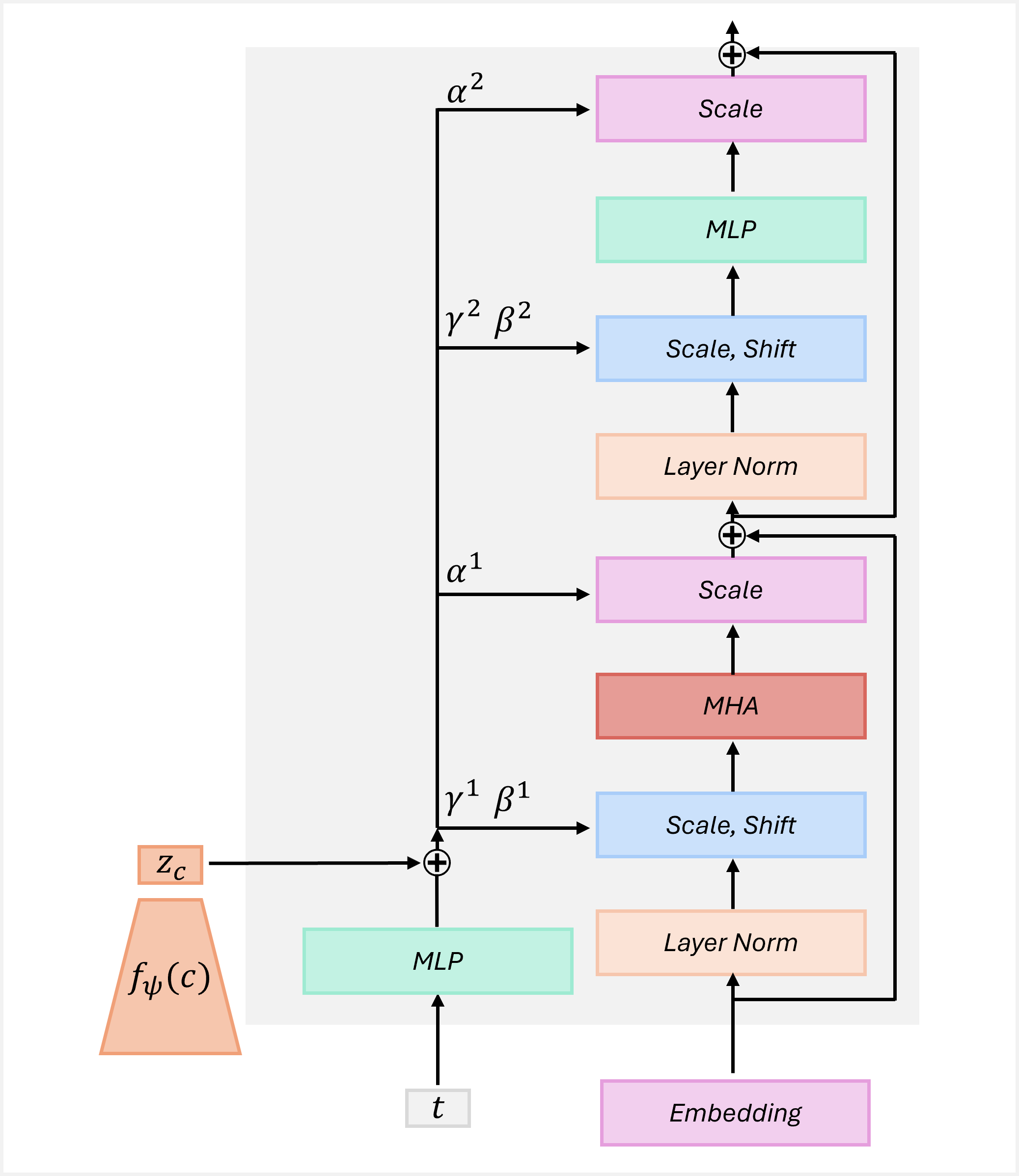}
            \caption{}
            \label{fig:subfig3}
        \end{subfigure}
        \hfill
        \begin{subfigure}{0.3\textwidth} 
            \centering
            \includegraphics[width=\linewidth]{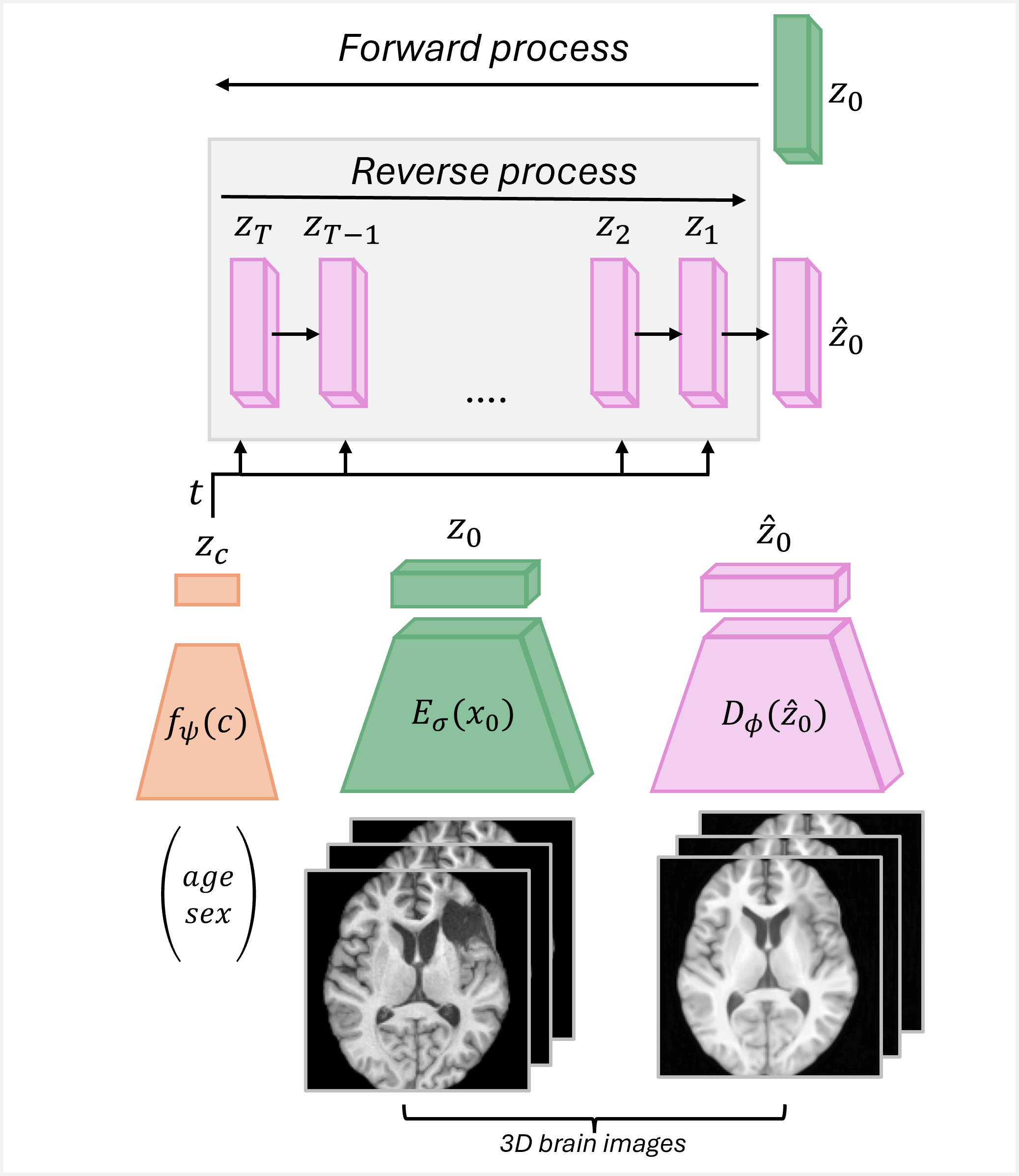}
            \caption{}
            \label{fig:subfig1}
        \end{subfigure}

    }
    \caption{Elements of the CADD model backbone. (a) The $\epsilon_\theta$ architecture, trained solely on healthy images. (b) The S-AdaLN architecture \cite{ma2024}, used for covariate conditioning. (c) An overview of the CADD diffusion model backbone applied at inference time to map from a disease input image to a pseudo-healthy reconstruction.}
    \label{fig:CADD_framework}
\end{figure*}
\section{Related work}
\noindent\textbf{Unsupervised medical anomaly detection.} Unsupervised anomaly detection has gained popularity in the medical field as it enables training solely on healthy images, removing the need for large disease cohorts or assumptions about anomaly characteristics \cite{zhou2020, pinaya2021, Jiang2023, schlegl2017, schlegl2019, zimmerer2018, chen2018, chen2020}. This is especially beneficial given the often limited availability of abnormal images.  One such paradigm for OoD are reconstruction-based methods which assume that a model trained on normal data cannot accurately represent or reconstruct anomalies \cite{Gong2019}. Autoencoders, which constitute a large portion of these methods \cite{Gong2019,zong2018,Baur2020,Denouden2018,Kumar2021}, involve training at least two mappings: an encoder which embeds information from the input space $X$ into a low-dimensional latent space $Z$, and a decoder which transforms samples from the latent space back into the input space. The latent space bottleneck limits the model's ability to faithfully reconstruct OoD, i.e. diseased, regions and will instead reconstruct a closely matched healthy counterpart. Abnormalities can then be detected by comparing the generated and original pathology images in pixel space. The generated images or subsequent anomaly maps have proven useful for a number of downstream tasks such as anomaly segmentation \cite{Baur2020}. However, several studies have found that autoencoders can accurately reconstruct various types of OoD samples \cite{Bercea2023,Zhou2023}, meaning that the resultant reconstruction errors do not fully capture the abnormality of the samples. Furthermore, autoencoder-based anomaly detection approaches suffer from poor generative ability, resulting in blurred reconstructions and high reconstruction errors, even for the healthy training distribution \cite{Bercea2023}.


\noindent\textbf{Diffusion models for medical anomaly detection.} DDPMs \cite{Ho2020} and DDIMs \cite{song2022} have demonstrated significant improvements in the effectiveness of anomaly detection of reconstruction-based methods \cite{Pinaya2022,Bercea2024c,Wolleb2022,Bercea2023b,Graham2023b,graham2023c,Iqbal2023}.  Diffusion models are able to capture complex data distributions by gradually adding and subsequently removing, typically gaussian, noise from an image in a set of noising and denoising steps respectively. In unsupervised anomaly detection, diffusion models function like other reconstruction-based methods by training on healthy images, now with the assumption that denoising a noised disease image will inpaint anomalous regions with healthy tissue. However, this process introduces a trade-off: selecting a noise level at inference time sufficiently high to remove anomalies risks erasing distinctive features of healthy tissue.

\noindent\textbf{Robust reconstruction of healthy tissue.} Several works have sought to address the trade-off between removing anomalies whilst retaining healthy tissue information. \cite{Graham2023b} choose to average reconstructions and similarity metrics across multiple noise levels. However this can result in blurry reconstructions. \cite{Bercea2023b}, instead combine partially noised images through masking, stitching, and re-sampling at various noise levels. Similarly, \cite{Bercea2024c} introduce THOR, a simplified scheme which uses an anomaly map for stitching across multiple noise levels. However, these methods involve complex partial noising, stitching and resampling procedures, or rely on image-space calculations that are impractical for 3D images. Additionally, both approaches use sample-wise metrics to identify healthy/unhealthy regions, which produce poor results for datasets containing healthy samples. In this work, we use a mask, generated from the KL-divergence between the model reverse and forward steps, to guide the denoising scheme and generate pseudo-healthy images. We modulate our mask using a KL-divergence threshold from a healthy holdout, ensuring that only regions which sit at the extremes of this distribution are inpainted. 

\noindent\textbf{Deep Normative modeling.} Without taking them into account, clinical covariates may manifest as confounders, variables which cause spurious associations or contribute to the causal pathway but are not of primary interest. Normative modeling addresses this by integrating covariates into the modeling framework. Typically, a normative analysis involves training a regression model, e.g., using Gaussian Process Regression (GPR)~\cite{Marquand:2016}, to predict a biomedical feature from a set of clinical covariates. However, traditional normative modeling approaches are computationally impractical for large 3D imaging datasets, do not consider the interactions between features, and lack generative capability. The ability to generate pseudo-healthy images can prove useful for downstream tasks \cite{durrer2024,Pinaya2022}. Recently, autoencoder \cite{pinaya2021,LawryAguila:2022,kumar2022,lawryaguila2023} and Transformer \cite{DaCosta2022} models have been proposed for normative modeling of neurological disorders. \cite{Ijishakin2024} present the first application of diffusion models to normative modeling, training a normative diffusion autoencoder on 2D brain images to predict survival in ALS using an encoder network. We instead introduce a normative diffusion model using a reconstruction-based anomaly detection approach to generate anomaly maps and abnormality indices for disease detection in the image space. 

\noindent\textbf{Diffusion Models with Transformers.} Recently, diffusion models using Transformers have been proposed \cite{Peebles2023} which outperform previously used U-net models and inherit good scaling properties from the Transformer model class. The application of Transformer models as diffusion model backbones has extended to the medical domain \cite{Wu2023,Chowdary2023} where it has been argued that there is some evidence to suggest that transformer-based models are better for capturing contextual information in medical images \cite{Chowdary2023}. In this work we use a Transformer backbone for a diffusion model trained to denoise noised healthy brain images.

\section{Methods}
\subsection{Preliminary of Latent Diffusion Models}

We consider LDMs for disease detection and image generation in brain images. LDMs are trained in two separate stages. The first stage model, uses an encoder $E_{\sigma}(\cdot)$ to compress the input image $\mathbf{x}_{0} \in \mathbb{R}^{H \times W \times D \times 1}$ to a lower-dimension latent representation $\mathbf{z}_{0} \in \mathbb{R}^{h \times w \times d \times c}$ and a decoder $D_{\phi}(\cdot)$ to map $\mathbf{z}_{0}$ back up to the input space. The second stage model is a Denoising Diffusion Probabilistic Model (DDPM)~\cite{Ho2020} trained to learn the distribution of our latent representation $\mathbf{z}_0 \sim q\left(\mathbf{z}_0\right)$. The diffusion model consists of two elements. (1) The \textit{forward process} which is a pre-defined Markov chain with $T$ gaussian transitions. This progressively noises the latent representation $\mathbf{z}_{0}$ such that $T$ noising steps approximates a prior distribution $\mathbf{z}_T \sim \mathcal{N}(0, \mathbf{I})$.  The forward process is defined as:
\begin{equation}
\begin{split}
    q(\mathbf{z}_{1:T}|\mathbf{z}_{0}) &:= \prod^{T}_{t=1} q(\mathbf{z}_{t}|\mathbf{z}_{t-1}),  \\
    q(\mathbf{z}_{t}|\mathbf{z}_{t-1}) &:= \mathcal{N}(\mathbf{z}_{t};\sqrt{\alpha_t} \mathbf{z}_{t-1}, \left(1-\alpha_t\right) \textbf{I})
\end{split}
\end{equation}
where $\mathbf{z}_t$ is the noisy latent feature sampled at diffusion timestep $t$, $t \in\{\mathbb{Z}|0\leq t \leq 1000\}$, and the parameter $\alpha_t \in \mathbb{R}$ controls the level of noise added at step $t$.

(2) The second element is the \textit{reverse process}. This is another Markov chain $p_\theta\left(\mathbf{z}_{0: T}\right) = p\left(\mathbf{z}_{T}\right)\prod_{t=1}^{T}p_{\theta}\left( \mathbf{z}_{t-1}|\mathbf{z}_{t}\right)$ that learns to recover the original data $\mathbf{z}_0$ from our prior $\mathbf{z}_T$. Each step is given by:
\begin{equation}
    p_\theta\left(\mathbf{z}_{t-1} \mid \mathbf{z}_t\right)=\mathcal{N}\left(\mu_\theta\left(\mathbf{z}_t\right), \Sigma_\theta\left(\mathbf{z}_t\right)\right).
\end{equation}
Given the forward and reverse processes we can construct a variational lower bound on the log-likelihood of our latent $\mathbf{z}_{0}$, which reduces to:
\begin{equation}
\begin{aligned}
    \label{eq:lower-bound}
    \mathbb{E} \left[ \log p_{\theta}(\mathbf{z}_{0}) \right] & \geq 
    \log p\left(\mathbf{z}_0 | \mathbf{z}_1\right) \\
    & - \sum_t D_{KL}\left(q\left(\mathbf{z}_{t-1} | \mathbf{z}_t, \mathbf{z}_0\right) \| p_\theta\left(\mathbf{z}_{t-1} | \mathbf{z}_t\right)\right).
\end{aligned}
\end{equation}
Following the parameterization from \cite{Ho2020}, $\mu_\theta$ can be modeled using a denoising model $\epsilon_\theta$ which can be trained with the simple objective:
\begin{equation}
\label{eq:obj}
\mathcal{L}=\mathbb{E}_{\mathbf{z}_{0} \sim q(\mathbf{z}_{0}), \epsilon_{t} \sim \mathcal{N}(\mathbf{0}, \mathbf{I})}\left[\left\|\epsilon_t-\epsilon_\theta\left(\mathbf{z}_t, t\right)\right\|_2^2\right]
\end{equation}

\subsection{CADD: Diffusion model framework}\label{sec:diffusion_model_network}
To incorporate contextual clinical information into our modeling framework, we must disentangle $\mathbf{z}_t$ from factors with known biological variance. Now let $\mathbf{z}_0 \sim q\left(\mathbf{z}_0|\mathbf{z}_c\right)$ where $\mathbf{z}_c$ is a representation of clinical covariates, $\boldsymbol{c}$. The diffusion model reverse process becomes 
$p_\theta\left(\mathbf{z}_{0: T}|\mathbf{z}_c\right) = p\left(\mathbf{z}_{T}\right)\prod_{t=1}^{T}p_{\theta}\left( \mathbf{z}_{t-1}|\mathbf{z}_{t}, \mathbf{z}_c\right)$ such that we now recover the original data distribution conditional on the clinical information, $\mathbf{z}_c$. It can be shown that Equation \ref{eq:obj} becomes:
\begin{equation}
\label{eq:obj_c}
\mathcal{L}=\mathbb{E}_{\mathbf{z}_{0} \sim q(\mathbf{z}_{0}|\mathbf{z}_c), \epsilon_{t} \sim \mathcal{N}(\mathbf{0}, \mathbf{I})}\left[\left\|\epsilon_t-\epsilon_\theta\left(\mathbf{z}_t,\mathbf{z}_c, t\right)\right\|_2^2\right].
\end{equation}
To learn $\mathbf{z}_{c}$, we train a network $f_{\psi}(\cdot)$ such that $(\mathbf{z}_c, \mathbf{z}_{t}) \in \mathbb{R}^{h}$ where $h$ represents the dimension of the hidden embedding in the Transformer blocks.

In this work, we implement $\epsilon_\theta$ as a Transformer \cite{ma2024}, as opposed to the more commonly used U-net \cite{Ronneberger2015,Rombach2021,Pinaya2022,Graham2023b}. To manage the computational burden of our model, we decompose our 3D latent embeddings into two components: $\mathbf{z}_{hw}$, which focuses on capturing relationships across the first two spatial dimensions, and $\mathbf{z}_{d}$, which captures information in the remaining spatial dimension. These components are processed through two distinct types of Transformer blocks. The alternating application of these Transformer blocks, illustrated in Figure \ref{fig:CADD_framework}, adopts an ``interleaved fusion" approach to effectively capture the relationship across all 3 dimensions (see Supp. for further details). For integrating timestep $t$ and covariate $\mathbf{z}_c$ information into our model, we apply the scalable adaptive layer normalization (S-AdaLN) proposed by \cite{ma2024} and shown in Figure \ref{fig:CADD_framework}.

\begin{algorithm}[t]
\SetAlgoLined
\caption{CADD Inference Inpainting Scheme}
\label{algo:inpaint}
\SetKwInOut{Input}{Input}
\SetKwInOut{Settings}{Settings}

\Input{$\mathbf{z}_0 = E(\mathbf{x}_0)$, $\mathbf{z}_c = f_{\psi}(\boldsymbol{c})$, $\text{KL}_{P_{95}}^{\text{val}}$ } 
\Settings{$T_\text{int}$}

\For{$U \in \{50 \cdot k : k = 1,\dots, \frac{T_{\text{int}}}{50}\}$}{
    \textbf{Generate} $\mathbf{z}_{U} \sim q(\mathbf{z}_{1:U} | \mathbf{z}_{0})$ \\
    \textbf{Calculate} $\text{KL}_{U} = D_{KL}\left(q\left(\mathbf{z}_{U-1} | \mathbf{z}_U, \mathbf{z}_0\right) \| p_\theta\left(\mathbf{z}_{U-1} | \mathbf{z}_U, \mathbf{z}_c\right)\right)$ \\

    $m_{\text{s}} = \begin{cases} 
        1 & \text{if } U_{\text{KL}} > P_{95}(\text{KL}_{U}) \\ 
        0 & \text{otherwise} 
    \end{cases}$ \\
    $m_{\text{v}} = \begin{cases} 
        1 & \text{if } U_{\text{KL}} > \text{KL}_{P_{95}}^{\text{val}}(U) \\ 
        0 & \text{otherwise} 
    \end{cases}$ \\
    $m = m_{\text{s}} \odot m_{\text{v}}$ \\[0.5em]

    \textbf{Generate } $ \mathbf{z}_{0}^{U}$ \\
    \Indp
        \For{$t = U,\, \ldots,\, 0$}{
            \textbf{Generate }$\mathbf{z}^{U}_{t-1} \sim p_\theta\left(\mathbf{z}_{t-1} | \mathbf{z}_t, \mathbf{z}_{c}\right)$
        }
        \Return $\mathbf{z}^{U}_{0}$ \\
    \Indm

    $\hat{\mathbf{z}}^{U}_0 = m \odot \mathbf{z}^{U}_0 + (1 - m) \odot \mathbf{z}_0$
}
\vspace{0.5em}
$\hat{\mathbf{z}}_0 = \frac{1}{N_U} \sum \hat{\mathbf{z}}^{U}_0$ \\
$\hat{\mathbf{x}}_0 = D(\hat{\mathbf{z}}_0)$
\end{algorithm}
\subsection{CADD: Inference inpainting scheme}

In diffusion model-based anomaly removal, the goal is to inpaint anomalous tissue while preserving healthy regions. This is done by applying a denoising model, $\epsilon_\theta$, trained on healthy brains, to gradually remove noise from a partially noised disease image. To address the anomaly removal vs individual characteristics preservation trade-off, we introduce an inpainting scheme inspired by \cite{Pinaya2022} where we selectively denoise anomalous regions using pixel-wise masks to guide the pseudo-healthy reconstruction process.

Consider the KL-divergence term in Equation \ref{eq:lower-bound}, at a given timestep $t$, we expect anomalous regions to deviate more greatly in the reverse process $p_\theta\left(\mathbf{z}_{t-1} | \mathbf{z}_t, \mathbf{z}_c\right)$ from the expected Gaussian transition $q\left(\mathbf{z}_{t-1} | \mathbf{z}_t, \mathbf{z}_0\right)$ than the healthy regions the model was trained on. To leverage this, we introduce Algorithm \ref{algo:inpaint}, which uses the KL-term to generate masks to guide the reconstruction process. We incorporate both sample-wise masks, $m_\text{s}$ (where $P_{95}(\text{KL}_{U})$ is calculated for each individual sample across the latent vector distribution), and vector-wise masks, $m_\text{v}$ (where $\text{KL}_{P_{95}}^{\text{val}}(U)$ is calculated for each vector across a validation cohort distribution). We use a 95\% abnormality threshold as in prior normative modeling work \cite{lawryaguila2023}. 
The pseudo-healthy reconstruction, generated from an intermediately timestep $T_\text{int}$, can be compared with the original image to generate anomaly maps, detect disease effects, or used for other downstream tasks. Whilst previous works \cite{Bercea2023b,Bercea2024c} require full image reconstruction or complex inpainting procedures, our method integrates masking into the denoising scheme for a more streamlined approach. 

\section{Experiments and Results}

\subsection{Experimental Setup}

\textbf{Datasets.} We evaluate our method on three medical datasets; The UK Biobank (UKBB) \cite{Sudlow:2015} (Application number 100955), the Alzheimer's Disease Neuroimaging Initiative (ADNI) \cite{Petersen:2010}, and our in-house University of Tokyo Hospital (UoTH) dataset. The UKBB dataset consists of healthy subjects used for model training (N=10,276), a healthy validation cohort (N=1070), a healthy test cohort (N=1070), and a disease cohort (N=122) with one of several neurodegenerative disorders; motor neuron disease, multiple sclerosis, Parkinson's disease, dementia/Alzheimer/cognitive-impairment and other demyelinating disease. Healthy subjects were selected such that they had no neurological, psychiatric disorders or head trauma. The ADNI dataset consists of a finetuning cohort (N=200), a healthy validation cohort (N=50), a healthy test cohort (N=54), and a disease cohort (N=180) of individuals with significant memory concern (SMC; N=52), early mild cognitive impairment (EMCI; N=89), late mild cognitive impairment (LMCI; N=37) and Alzheimer's disease (AD; N=147). The UoTH dataset consists of a finetuning cohort (N=269), validation cohort (N=32), test cohort (N=32) and a disease cohort (N=58) of individuals with gliomas (N=42) and infarcts (N=16). For the ADNI and UoTH datasets the finetuning cohort is used to finetune models pre-trained on the UKBB healthy training cohort. For all datasets the validation cohorts are used for early stopping and to generate z-score metrics. We use T1-weighted MRI scans which all underwent the same pre-processing steps (described in the Supp.) resulting in a dimensionality of $128 \times 128 \times 128$.

\noindent\textbf{Comparison methods.} We compare our model to the following generative modeling based anomaly-detection approaches; a VAE \cite{Baur2020}, cVAE \cite{LawryAguila:2022}, LDM \cite{Rombach2021}, LDM ($T_{\text{avg}}$) \cite{Graham2023b}, THOR \cite{Bercea2024c} and AutoDDPM \cite{Bercea2023b}. Where possible and available, we use the code from the original implementation. For the cVAE, we use the VAE CNN encoder and decoder architecture \cite{Baur2020} and condition on age and sex by projecting these covariates as extra channels of the encoder input and concatenating to the latent space for the decoder input. We use the same first stage model for LDM baselines and CADD. Since the main contribution of THOR and AutoDDPM lies in their inpainting schemes, we use a transformer backbone for the DDPM for closer comparison to our work. As image-space DDPM is computationally infeasible for 3D brain images, we implement both methods as LDMs and extend their 2D inpainting schemes to 3D. We term these adaptations of the original methods THOR (3D) and AutoDDPM (3D). See Supp. for further details on how we adapt THOR and AutoDDPM to 3D brain images.

\noindent\textbf{Implementation details.} As the first stage, we use an Autoencoder with a KL-regularised latent space and perceptual and patch-based adversarial objectives \cite{Rombach2021} which maps the brain image to a latent representation of size $3 \times 16 \times 16 \times 16$. We use the training parameters given by \cite{pinaya2022b}. For the second stage, CADD uses the Transformer backbone (see Section \ref{sec:diffusion_model_network}). We use 28 transformer blocks, each with 16 attention heads and a latent size of 1024 for each attention head. Timesteps are sinusoidally embedded and processed through a two-layer MLP with Swish activation \cite{ramachandran2017}, resulting in a 1024-dimensional embedding. Clinical covariates, specifically age and sex, are incorporated by passing them through $f_{\psi}(\cdot)$, implemented as a single-layer MLP. This timestep and covariate information is integrated into the model via S-AdaLN modules in each transformer block. During training, we use $T = 1000$ and apply a linear noise schedule with $\beta_t$ ranging from 0.0015 to 0.0195. All models are trained using the Adam optimiser \cite{kingma2017} with an early stopping criteria on the validation loss and a learning rate of 0.0001. For the UKBB and ADNI datasets we use $T_\text{int}=250$ at inference time for anomaly detection, inspired by prior works \cite{Bercea2023b,Bercea2024c}. For fair comparison, we use the same $T_\text{int}$ for the LDM baseline. For the UoTH dataset, as we expect the noisier clinical images to require further inpainting steps, we use the full noising chain. See Supp. Section \ref{sec:Timestep_analysis} for an analysis of noise level. All models use a random seed of 42 for inference and data folds.

\begin{table*}[hbt!]
    \centering
    \smaller
    \setlength{\abovecaptionskip}{3pt} 
    \setlength{\belowcaptionskip}{0pt} 
    \begin{tabularx}{\textwidth}{p{0.5cm}p{2.4cm}XXXp{2cm}p{2cm}p{2.1cm}}
        \hline
        Dataset & Method &  MAE ($\downarrow$) & PSNR ($\uparrow$) & $\operatorname{SSIM}$ ($\uparrow$) & $\text{LPIPS}_{\text{alex}}$ ($\downarrow$) & $\text{LPIPS}_{\text{vgg}}$ ($\downarrow$) & $\text{LPIPS}_{\text{squeeze}}$ ($\downarrow$) \\
        \hline
        \multirow{8}{*}{UKBB} & VAE \cite{Baur2020} & 0.0234$\pm$0.0003 & 26.2386$\pm$0.1147 & 0.8205$\pm$0.0009 & 0.1983$\pm$0.0007 & 0.2375$\pm$0.0008 & 0.1569$\pm$0.0004 \\
        & cVAE \cite{LawryAguila:2022} & 0.0205$\pm$0.0003 & 27.0139$\pm$0.1163 & 0.8525$\pm$0.0007 & 0.1785$\pm$0.0006 & 0.1787$\pm$0.0003 & 0.1373$\pm$0.0004 \\
        & LDM & 0.0332$\pm$0.0004 & 22.5156$\pm$0.1113 & 0.7650$\pm$0.0006 & 0.0942$\pm$0.0003 & 0.1506$\pm$0.0003 & 0.0764$\pm$0.0003 \\ 
        & LDM ($T_{\text{avg}}$) \cite{Graham2023b} & 0.0441$\pm$0.0004 & 20.8605$\pm$0.0969 & 0.7299$\pm$0.0006 & 0.1152$\pm$0.0004 & 0.1743$\pm$0.0004 & 0.0931$\pm$0.0003 \\
        & AutoDDPM (3D) \cite{Bercea2023b} & 0.0219$\pm$0.0003 & 25.4312$\pm$0.1177 & 0.8712$\pm$0.0007 & 0.0762$\pm$0.0004 & 0.1196$\pm$0.0005 & 0.0639$\pm$0.0003 \\
        & THOR (3D) \cite{Bercea2024c} & \underline{0.0114$\pm$0.0002} & \underline{31.9337$\pm$0.1772} & \underline{0.9503$\pm$0.0007} & \underline{0.0582$\pm$0.0008} & \underline{0.0838$\pm$0.0006} & \underline{0.0523$\pm$0.0005} \\
        & CADD (Ours) & \textbf{0.0103$\pm$0.0001} & \textbf{32.1909$\pm$0.1206} & \textbf{0.9543$\pm$0.0003} & \textbf{0.0406$\pm$0.0003} & \textbf{0.0740$\pm$0.0003} & \textbf{0.0404$\pm$0.0003} \\
        \hline
        \multirow{8}{*}{ADNI} & VAE \cite{Baur2020} & 0.0391$\pm$0.0014 & 21.7259$\pm$0.3311 & 0.7662$\pm$0.0073 & 0.1648$\pm$0.0027 & 0.2517$\pm$0.0053 & 0.1316$\pm$0.0025 \\ 
        & cVAE \cite{LawryAguila:2022} & 0.0331$\pm$0.0012 & 22.6978$\pm$0.3320 & 0.8308$\pm$0.0030 & 0.1839$\pm$0.0031 & 0.1868$\pm$0.0026 & 0.1348$\pm$0.0022 \\
        & LDM & 0.0489$\pm$0.0014 & 19.1645$\pm$0.2923 & 0.7489$\pm$0.0032 & 0.1054$\pm$0.0021 & 0.1587$\pm$0.0018 & 0.0799$\pm$0.0015 \\ 
        & LDM ($T_{\text{avg}}$) \cite{Graham2023b} & 0.0589$\pm$0.0014 & 17.9214$\pm$0.2433 & 0.7134$\pm$0.0024 & 0.1250$\pm$0.0019 & 0.1849$\pm$0.0020 & 0.0961$\pm$0.0017 \\ 
        & AutoDDPM (3D) \cite{Bercea2023b} & 0.0330$\pm$0.0009 & 21.7522$\pm$0.2559 & 0.8569$\pm$0.0036 & 0.0815$\pm$0.0020 & 0.1302$\pm$0.0022 & 0.0659$\pm$0.0016 \\  
        & THOR (3D) \cite{Bercea2024c} & \underline{0.0213$\pm$0.0010} & \underline{26.3221$\pm$0.4942} & \underline{0.9232$\pm$0.0022} & \underline{0.0789$\pm$0.0029} & \underline{0.0957$\pm$0.0019} & \underline{0.0627$\pm$0.0018} \\
        & CADD (Ours) & \textbf{0.0162$\pm$0.0006} & \textbf{28.0765$\pm$0.3532} & \textbf{0.9486$\pm$0.0027} & \textbf{0.0425$\pm$0.0016} & \textbf{0.0797$\pm$0.0025} & \textbf{0.0414$\pm$0.0017} \\
        \hline
        \multirow{8}{*}{UoTH} & VAE \cite{Baur2020} & 0.0259$\pm$0.0013 & 25.6269$\pm$0.4420 & 0.7514$\pm$0.0051 & 0.2175$\pm$0.0051 & 0.2802$\pm$0.0038 & 0.1755$\pm$0.0043 \\        
        & cVAE \cite{LawryAguila:2022} & 0.0215$\pm$0.0011 & 26.3920$\pm$0.4417 & 0.8437$\pm$0.0047 & 0.1866$\pm$0.0045 & 0.1978$\pm$0.0043 & 0.1443$\pm$0.0039 \\
        & LDM & 0.0380$\pm$0.0022 & 21.4159$\pm$0.5125 & 0.7392$\pm$0.0098 & 0.1153$\pm$0.0042 & 0.1789$\pm$0.0047 & 0.1037$\pm$0.0037 \\
        & LDM ($T_{\text{avg}}$) \cite{Graham2023b} & 0.0484$\pm$0.0011 & 19.9405$\pm$0.1544 & 0.7041$\pm$0.0035 & 0.1400$\pm$0.0026 & 0.2052$\pm$0.0028 & 0.1235$\pm$0.0022 \\
        & AutoDDPM (3D) \cite{Bercea2023b} & \textbf{0.0148$\pm$0.0010} & \textbf{28.7499$\pm$0.5843} & \underline{0.9187$\pm$0.0047} & \textbf{0.0646$\pm$0.0027} & \underline{0.1145$\pm$0.0034} & \underline{0.0686$\pm$0.0028} \\
        & THOR (3D) \cite{Bercea2024c} & 0.0164$\pm$0.0014 & 28.4717$\pm$0.7527 & 0.9170$\pm$0.0058 & 0.0853$\pm$0.0035 & 0.1185$\pm$0.0036 & 0.0806$\pm$0.0031 \\
        & CADD (Ours) & \underline{0.0152$\pm$0.0009} & \underline{28.4761$\pm$0.6728} & \textbf{0.9193$\pm$0.0050} & \underline{0.0654$\pm$0.0033} & \textbf{0.1135$\pm$0.0039} & \textbf{0.0682$\pm$0.0031} \\
        \hline
    \end{tabularx}
    \caption{Image quality evaluation metrics and 95\% confidence intervals for CADD and baseline methods. For each dataset and metric, \textbf{
    bold} indicates the best results, and \underline{underlined} indicates the second best performance.}
    \label{tab:image_quality_results}
\end{table*}

\noindent\textbf{Comparison metrics.} Ideally, a model should correctly identify disease individuals, or individuals with anomalous regions, as outliers and healthy individuals as sitting within the normative distribution. Furthermore, for the generative models, we want a model which can generate high quality, realistic pseudo-healthy reconstructions for downstream tasks. As such, we assess the performance of our model against three tasks ability to; generate high quality images, detect anomalies agnostic of the particular disease, and detect disease specific effects. Methods which perform well against all three tasks can be considered to have addressed the anomaly detection vs healthy context trade-off whilst also effectively encoding disease-related information. 

\begin{table*}[h!]
    \centering
    \smaller
    \setlength{\abovecaptionskip}{3pt} 
    \setlength{\belowcaptionskip}{0pt} 
    \begin{tabularx}{\textwidth}{lp{2.4cm}XXXXXXXX}
        \hline
         \multicolumn{2}{c}{z-score type:}  & \multicolumn{2}{c}{MAE (1\%)} & \multicolumn{2}{c}{MAE (5\%)} & \multicolumn{2}{c}{MAE*LPIPS (1\%)} & \multicolumn{2}{c}{MAE*LPIPS (5\%)} \\
        Dataset & Method & AUC & p-value & AUC  & p-value & AUC  & p-value  & AUC & p-value  \\
        \hline
        \multirow{8}{*}{UKBB} & VAE \cite{Baur2020} & 0.5841 & 3.597E-6 & 0.5743 & 7.766E-5 & 0.5779 & 1.911E-5 & 0.5690 & 2.475E-4 \\ 
        & cVAE \cite{LawryAguila:2022} & \underline{0.5903} & \underline{3.180E-7} & \underline{0.5851} & \underline{3.267E-6} & \underline{0.5884} & \underline{9.273E-7} & \underline{0.5826} & \underline{8.273E-6} \\
        &LDM & 0.5464 & 8.029E-3 & 0.5312 & 9.450E-2 & 0.5485 & 7.528E-3 & 0.5353 & 6.296E-2 \\ 
        & LDM ($T_{\text{avg}}$) \cite{Graham2023b} & 0.5397 & 3.603E-2 & 0.5292 & 1.591E-1 & 0.5410 & 3.792E-2 & 0.5310 & 1.468E-1 \\ 
        & AutoDDPM (3D) \cite{Bercea2023b} & 0.5719 & 8.940E-5 & 0.5760 & 3.415E-5 & 0.5647 & 4.770E-4 & 0.5681 & 2.598E-4 \\ 
        & THOR (3D) \cite{Bercea2024c} & 0.5831 & 6.986E-6 & 0.5777 & 2.978E-5 & 0.5698 & 3.820E-4 & 0.5658 & 1.275E-3 \\
        & CADD (Ours) & \textbf{0.6052} & \textbf{6.828E-9} & \textbf{0.6017} & \textbf{2.586E-8} & \textbf{0.5994} & \textbf{5.585E-8} & \textbf{0.5967} & \textbf{1.771E-7} \\
        \hline
        \multirow{8}{*}{ADNI} 
        & VAE \cite{Baur2020} & \underline{0.5795} & \underline{1.200E-2} & \underline{0.5841} & \underline{2.480E-2} & \underline{0.6256} & \underline{1.743E-4} & \underline{0.6263} & \underline{4.208E-4} \\
        & cVAE \cite{LawryAguila:2022} & 0.5582 & 3.962E-2 & 0.5611 & 5.518E-2 & 0.5958 & 2.293E-3 & 0.6011 & 3.068E-3 \\
        & LDM & 0.5387 & 2.301E-1 & 0.5394 & 2.796E-1 & 0.5696 & 1.339E-1 & 0.5697 & 1.314E-1 \\ 
        & LDM ($T_{\text{avg}}$) \cite{Graham2023b} & 0.5307 & 3.616E-1 & 0.5279 & 4.231E-1 & 0.5694 & 6.812E-2 & 0.5710 & 7.324E-2 \\ 
        & AutoDDPM (3D) \cite{Bercea2023b} & 0.5792 & 5.528E-2 & 0.5720 & 3.761E-1 & 0.6106 & 4.071E-3 & 0.6002 & 9.172E-3 \\
        & THOR (3D) \cite{Bercea2024c} & 0.5627 & 1.360E-2 & 0.5678 & 2.570E-2 & 0.5903 & 2.893E-3 & 0.5919 & 4.072E-3 \\ 
        &CADD (Ours) & \textbf{0.5847} & \textbf{2.000E-3} & \textbf{0.5962} & \textbf{3.000E-3} & \textbf{0.6412} & \textbf{8.057E-5} & \textbf{0.6408} & \textbf{1.772E-4} \\ 
        \hline
        \multirow{8}{*}{UoTH} & VAE \cite{Baur2020} & \underline{0.8056} & \underline{8.770E-7} & 0.7456 & 9.310E-5 & 0.8013 & \underline{2.061E-6} & 0.7394 & 1.970E-4 \\ 
        & cVAE \cite{LawryAguila:2022} & \textbf{0.8069} & \textbf{6.231E-7} & \textbf{0.7863} & \textbf{1.924E-6} & \textbf{0.8288} & \textbf{9.034E-8} & \textbf{0.8125} & \textbf{1.815E-7} \\ 
        & LDM & 0.5775 & 2.780E-1 & 0.5288 & 7.613E-1 & 0.6050 & 3.078E-2 & 0.5581 & 1.129E-1 \\ 
        & LDM ($T_{\text{avg}}$) \cite{Graham2023b} & 0.5063 & 6.716E-1 & 0.4413 & 2.655E-1 & 0.5344 & 6.450E-1 & 0.4975 & 8.711E-1 \\ 
        & AutoDDPM (3D) \cite{Bercea2023b} & 0.7350 & 2.182E-4 & 0.7038 & 1.140E-3 & 0.7688 & 3.369E-5 & 0.7494 & 1.110E-4 \\ 
        & THOR (3D) \cite{Bercea2024c} & 0.6281 & 4.047E-2 & 0.5931 & 1.747E-1 & 0.6138 & 1.175E-1 & 0.5731 & 3.257E-1 \\ 
        & CADD (Ours) & 0.7631 & 3.688E-5 & \underline{0.7525} & \underline{2.174E-5} & \underline{0.8056} & 3.235E-6 & \underline{0.7881} & \underline{1.235E-6} \\ 
        \hline
    \end{tabularx}
    \caption{Disease detection evaluation of CADD and baseline methods. For each dataset and metric, \textbf{bold} indicates the best results, and \underline{underlined} indicates the second best performance.}
    \label{tab:disease_detection_results}
\end{table*}

\begin{table*}[hbt!]
    \centering
    \smaller
    \setlength{\abovecaptionskip}{3pt} 
    \setlength{\belowcaptionskip}{0pt} 
    \begin{tabularx}{\textwidth}{p{3.2cm}XXXXp{2.2cm}p{1.6cm}p{1.6cm}}
        \hline
        Rank ($\downarrow$) & VAE & cVAE & LDM & LDM ($T_{\text{avg}}$) & AutoDDPM (3D) & THOR (3D) & CADD (Ours) \\
        \hline
        Image quality (Table \ref{tab:image_quality_results}) & 5.83 & 4.89 & 5.00 & 6.06 & 2.72 & \underline{2.33} & \textbf{1.17} \\
        Disease detection (Table \ref{tab:disease_detection_results}) & 2.79 &	\underline{2.29} & 6.04 & 6.83 & 4.17 & 4.41 & \textbf{1.45} \\
        \hline
    \end{tabularx}
    \caption{Overall ranks for the image quality and disease detection tasks. Overall rank is calculated as the average across the ranks for each metric and dataset comparison for each task.}
    \label{tab:ranks}
\end{table*}
For image quality evaluation, we use average mean absolute error (MAE), peak signal-to-noise ratio (PSNR), structural similarity (SSIM) \cite{Wang2004}, and learned perceptual image patch similarity (LPIPS) \cite{Zhang2018} with AlexNet \cite{Krizhevsky2012}, VGGNet \cite{Simonyan2015}, and SqueezeNet \cite{Forrest2016} backbones. As the LPIPS metrics are designed for 2D images, we adopt a 2.5D approach. We calculate these metrics for the healthy holdout cohorts for each dataset to assess the ability of our proposed model to effectively reconstruct healthy tissue.

For disease detection, we calculate pixel-wise MAE and MAE*$\text{LPIPS}_{\text{Alex}}$ (weighting the pixel-wise metric by whole image similarity) for the disease cohort and healthy holdout cohort of each dataset. For each measurement, we generate z-scores using the measurements from the healthy validation cohort and use extreme value statistics \cite{Marquand:2016} (using a top 1\% and 5\% threshold) to aggregate abnormality across pixels and generate a single subject-level abnormality index. We calculate AUC scores and conduct Welch t-tests between the healthy and disease cohorts using each derived abnormality z-score metric. To generate abnormality maps, we calculate the pixel-wise MAE between the original image and its pseudo-healthy reconstruction.

For the ADNI dataset, we assess our model's ability to encode disease-related information by examining its sensitivity to patient cognition. We report the Pearson Correlation Coefficient, $\rho$, between z-score MAE (1\%) and age-adjusted memory, executive function, language, and visuospatial functioning composite scores for a subset (N=198) with available scores.

\begin{figure*}
    \centering
    \setlength{\abovecaptionskip}{3pt} 
    \setlength{\belowcaptionskip}{0pt} 
    \begin{subfigure}{0.06\linewidth} 
        \centering
        \includegraphics[width=\linewidth, trim=5 5 666 30, clip]{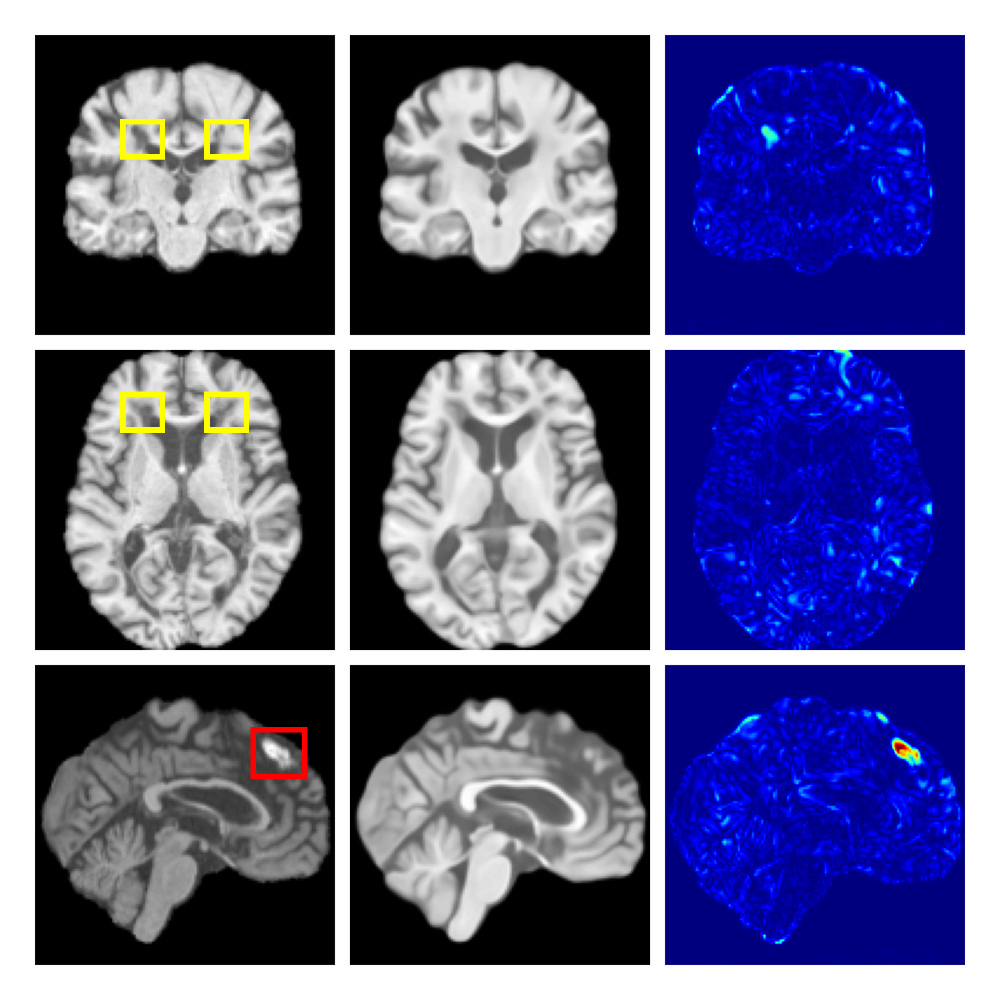}
        \caption{Input}
    \end{subfigure}
    \begin{subfigure}{0.12\linewidth} 
        \centering
        \includegraphics[width=\linewidth, trim=245 5 5 30, clip]{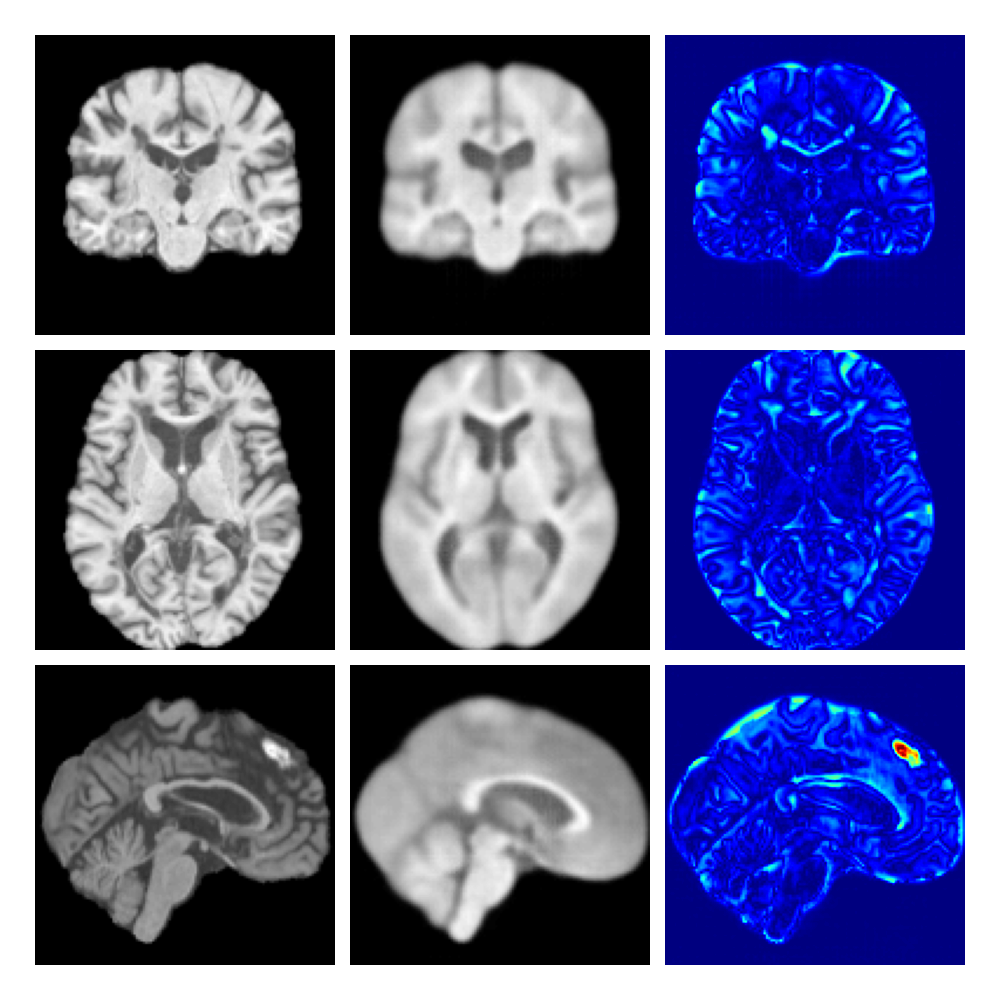}
        \caption{VAE}
    \end{subfigure}
    \begin{subfigure}{0.12\linewidth}
        \centering
        \includegraphics[width=\linewidth, trim=245 5 5 30, clip]{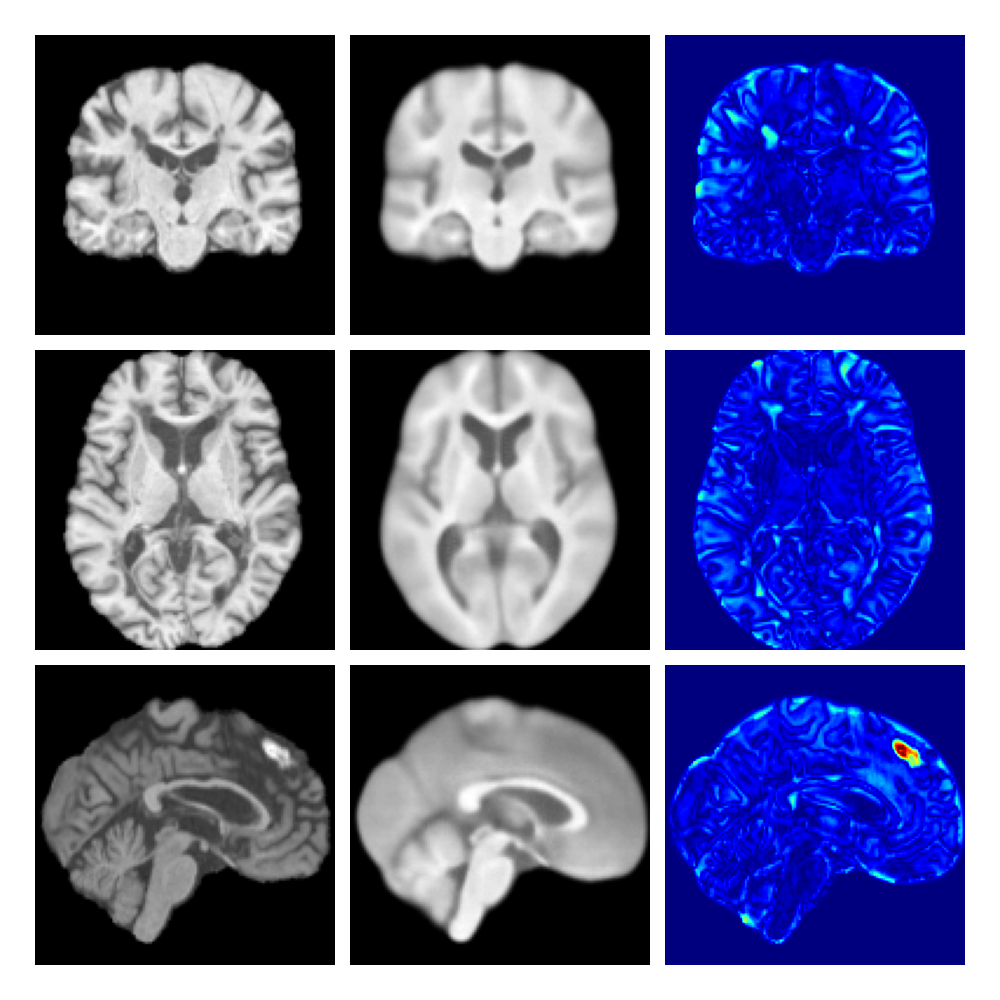}
        \caption{cVAE}
    \end{subfigure}
    \begin{subfigure}{0.12\linewidth}
        \centering
        \includegraphics[width=\linewidth, trim=245 5 5 30, clip]{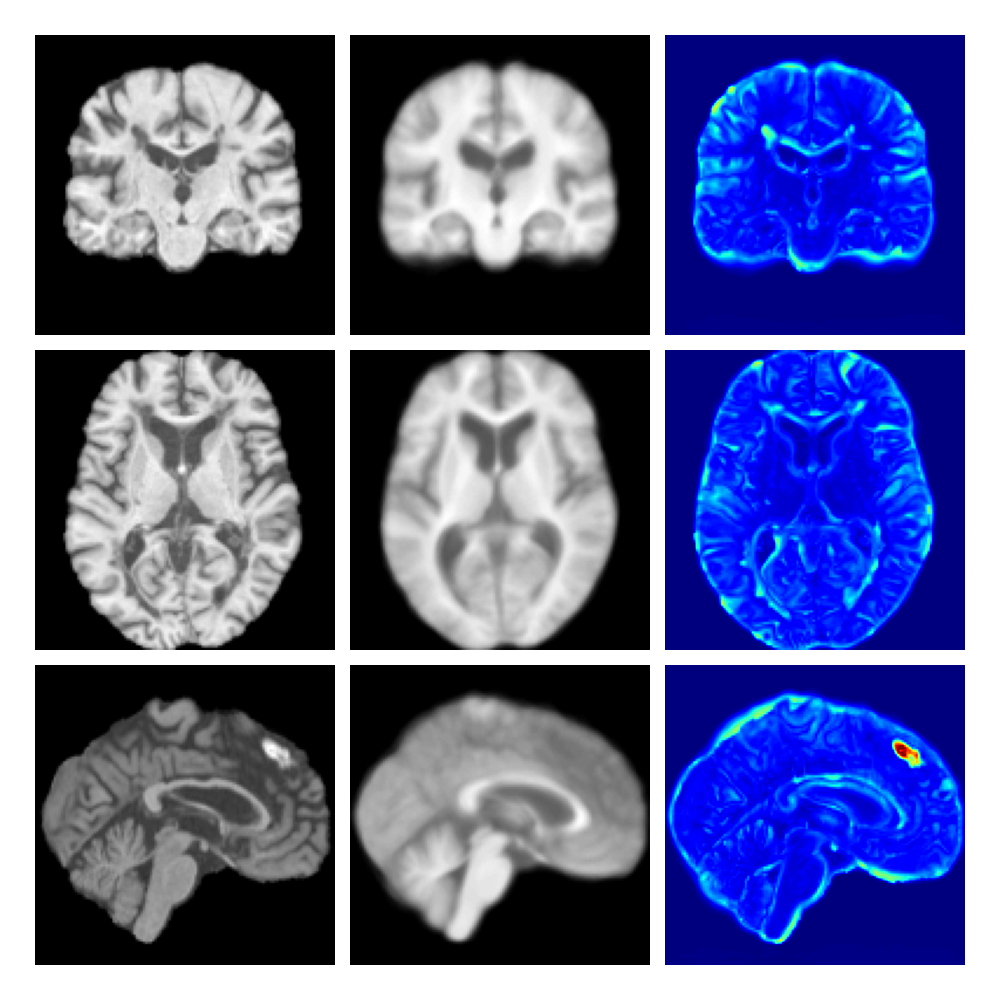}
        \caption{LDM ($T_{\text{avg}}$)}
    \end{subfigure}
    \begin{subfigure}{0.12\linewidth}
        \centering
        \includegraphics[width=\linewidth, trim=245 5 5 30, clip]{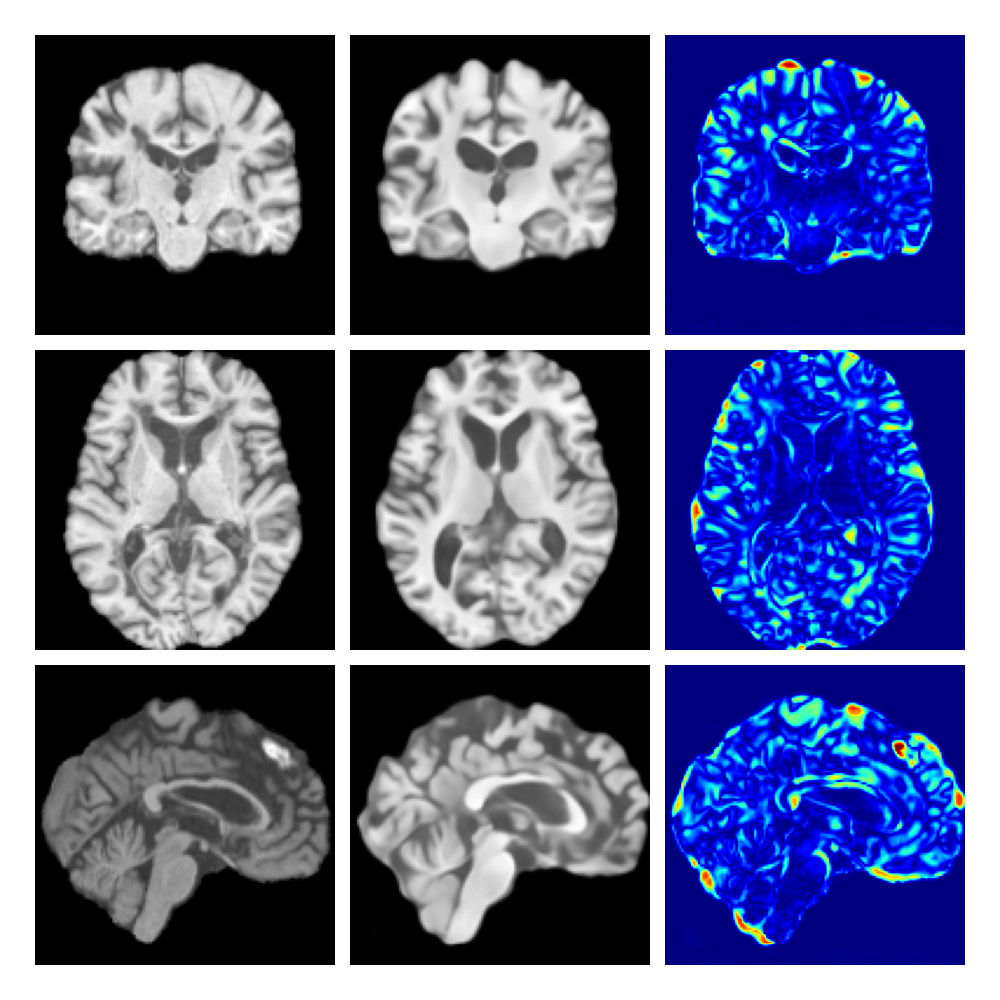}
        \caption{LDM}
    \end{subfigure}
    \begin{subfigure}{0.12\linewidth}
        \centering
        \includegraphics[width=\linewidth, trim=245 5 5 30, clip]{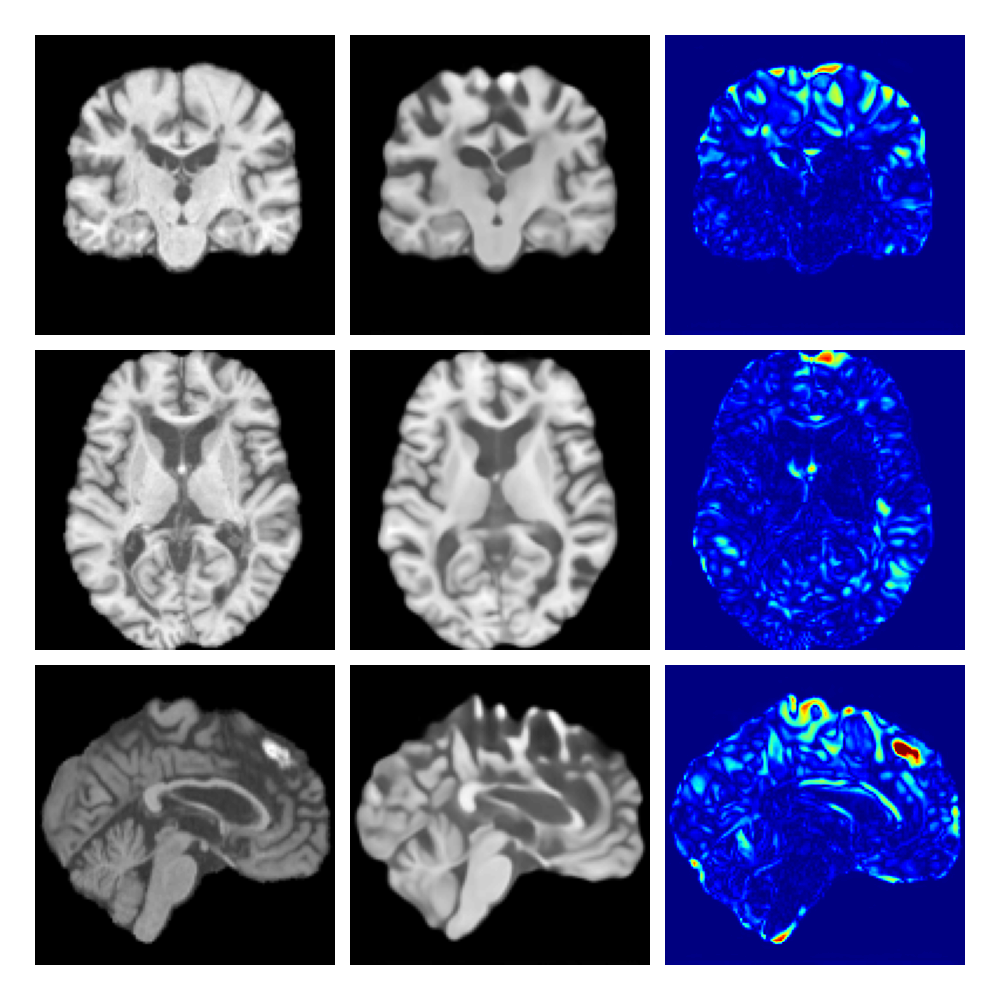}
        \caption{AutoDDPM}
    \end{subfigure}
    \begin{subfigure}{0.12\linewidth}
        \centering
        \includegraphics[width=\linewidth, trim=245 5 5 30, clip]{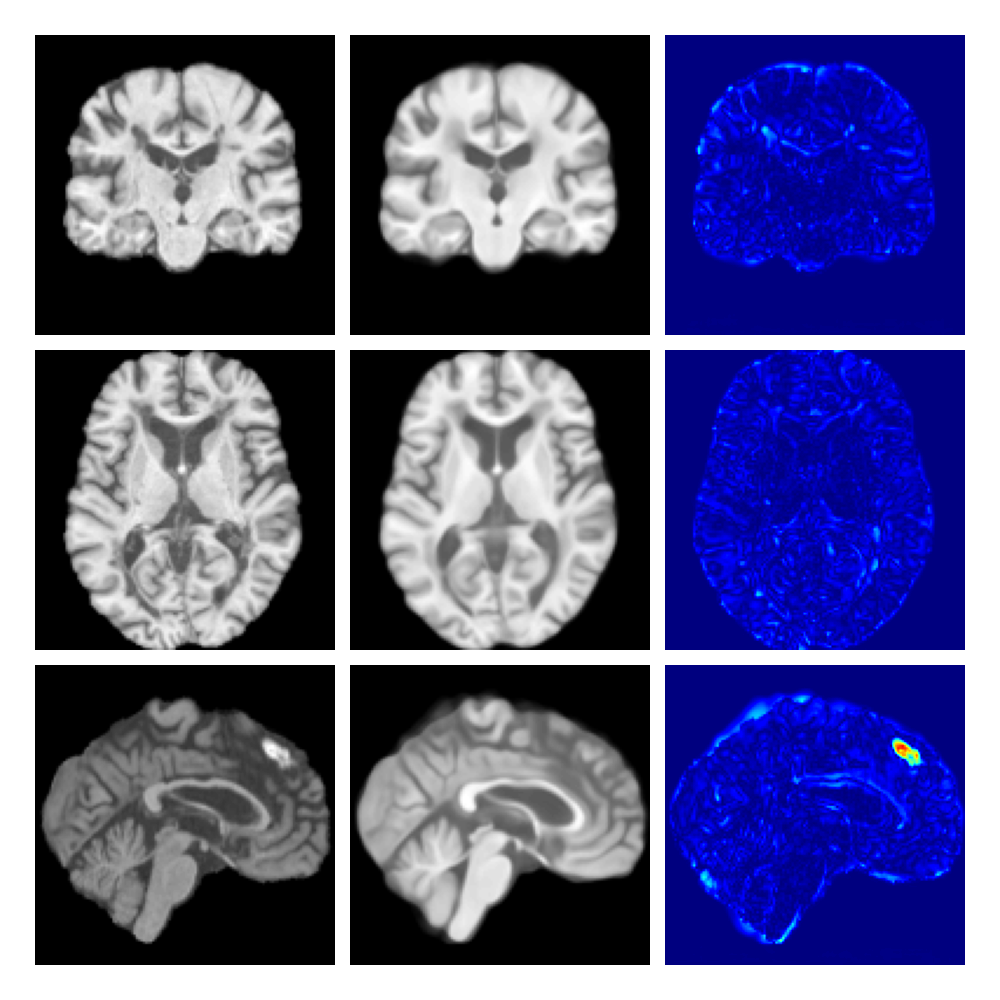}
        \caption{THOR}
    \end{subfigure}
    \begin{subfigure}{0.12\linewidth}
        \centering
        \includegraphics[width=\linewidth, trim=340 5 5 30, clip]{figures/ukbb/disease/5938271_CADD_thresh06.png}
        \caption{CADD (Ours)}
    \end{subfigure}
    
    \caption{Example reconstructions and anomaly maps for a sample from the disease cohort of the UKBB dataset. Lesion and WMH are indicated in the original image by the red and yellow boxes respectively.}
    \label{fig:example_UKBB_results}
\end{figure*}

\begin{figure}[hbt!]
    \centering
    \setlength{\abovecaptionskip}{3pt} 
    \setlength{\belowcaptionskip}{0pt} 
    \begin{subfigure}{0.125\columnwidth} 
        \centering
        \includegraphics[width=\linewidth, trim=5 5 666 30, clip]{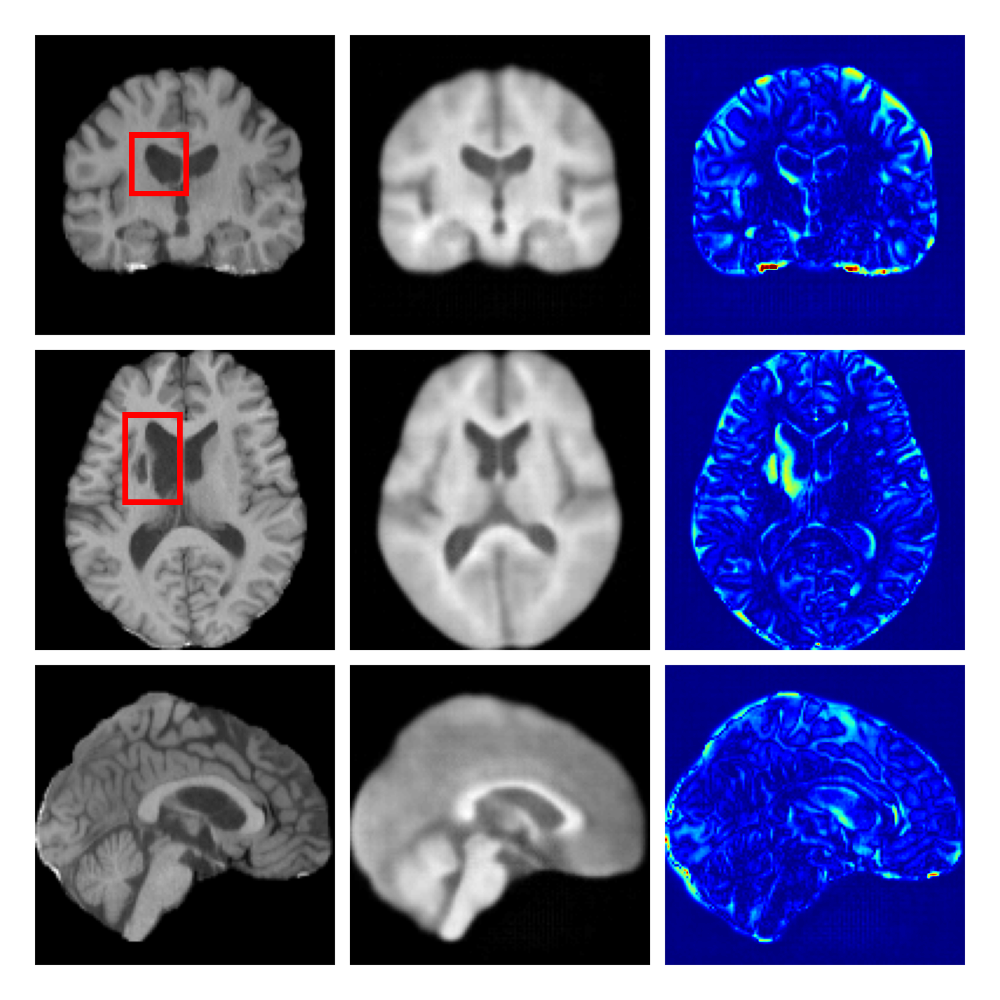}
        \caption{Input}
    \end{subfigure}
    \begin{subfigure}{0.249\columnwidth} 
        \centering
        \includegraphics[width=\linewidth, trim=340 5 5 30, clip]{figures/XXXH/disease/XF2F8071B19061DB9_VAE_thresh05.png}
        \caption{VAE}
    \end{subfigure}
    \begin{subfigure}{0.25\columnwidth}
        \centering
        \includegraphics[width=\linewidth, trim=245 5 5 30, clip]{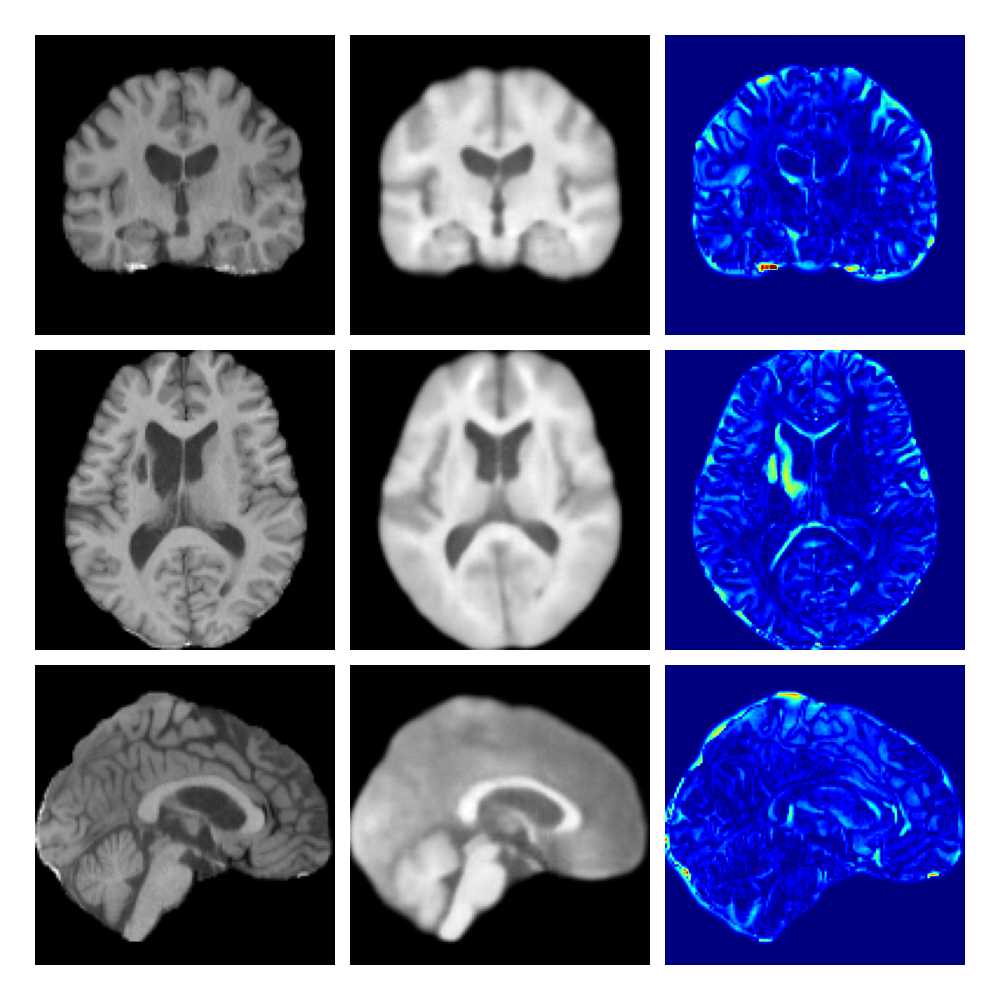}
        \caption{cVAE}
    \end{subfigure}
    \begin{subfigure}{0.25\columnwidth}
        \centering
        \includegraphics[width=\linewidth, trim=245 5 5 30, clip]{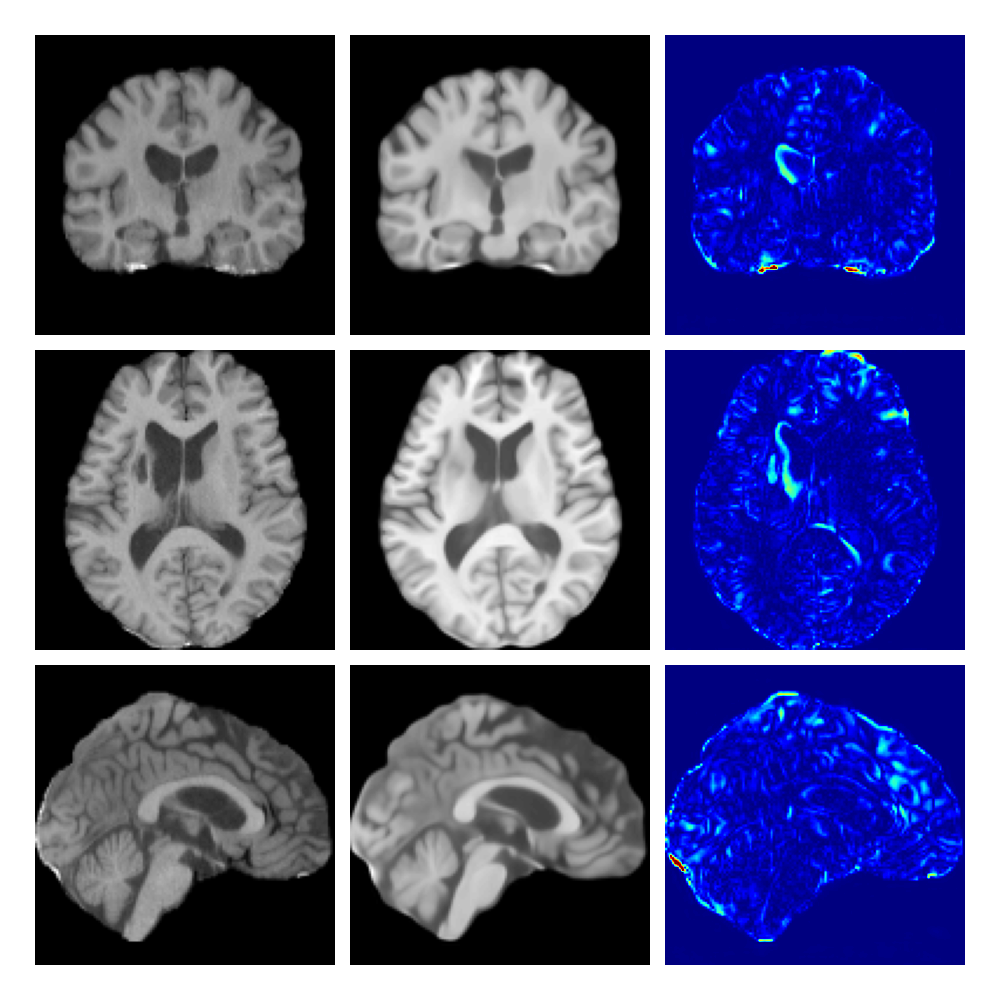}
        \caption{CADD (Ours)}
    \end{subfigure}
    
    \caption{Example reconstructions and anomaly maps from a disease cohort sample in the UoTH dataset, shown for the top three models. The lesion is highlighted in red in the original image.}
    \label{fig:example_UoT_results}
\end{figure}

\subsection{Results}

\noindent\textbf{Image quality evaluation.} Table \ref{tab:image_quality_results} (and Supp. Figure \ref{fig:example_UKBB_healthy}) show the healthy cohort image reconstruction results. CADD achieves the best (16 out of 18 tasks), or second best (2 out of 18 tasks), performance across all metrics and datasets with the best overall rank (Table \ref{tab:ranks}), followed by THOR (3D). This is expected, as CADD, THOR (3D), and also AutoDDPM (3D), incorporate some portion of $\textbf{z}_0$ into the reconstruction through their inpainting procedures. AutoDDPM (3D) demonstrates weaker performance for the UKBB and ADNI datasets, likely due to its indirect use of elements from $\textbf{z}_0$. Instead, it reconstructs from a partially noised, stitched $\textbf{z}$ through multiple denoising steps at a low $T$ value, which results in information loss. THOR (3D) underperforms compared to CADD, possibly because its sample-normalized anomaly masks may incorrectly inpaint healthy regions in a healthy cohort (e.g. see Supp. Figure \ref{fig:example_UKBB_healthy}). In contrast, CADD modulates the sample-wise mask with a vector-wise mask from the healthy holdout cohort, ensuring only regions at the extremes of this distribution are inpainted. AutoDDPM (3D) performs better on the UoTH dataset, where noisier images may benefit from the multiple resampling steps in the AutoDDPM inpainting scheme. As expected, the VAE and cVAE perform poorly in the image quality task as illustrated by blurry reconstructions in Figure \ref{fig:example_UKBB_healthy} and poor performance for the image quality metrics (Tables \ref{tab:image_quality_results} and \ref{tab:ranks}). For all datasets, we observe improved performance for the cVAE compared to VAE, suggesting that by incorporating contextual information we are able to better guide the reconstruction of healthy tissue.



\noindent\textbf{Disease detection evaluation.} Table \ref{tab:disease_detection_results} presents quantitative disease detection results. CADD performs consistently across all metrics and datasets with the best overall rank (Table \ref{tab:ranks}). Interestingly, while the cVAE outperforms the VAE on the UKBB and UoTH datasets, it underperforms on ADNI. Age is a known confounder of AD \cite{Liu2024}, and if not accounted for, could inflate pathological effects. Whether this effect is desirable depends on whether one wishes to consider healthy ageing effects anomalous or if the objective is to detect AD whilst taking into consideration expected changes from healthy ageing.

Figures \ref{fig:example_UKBB_results} and \ref{fig:example_UoT_results}, show example reconstructions and abnormality maps for a sample from the UKBB and UoTH datasets, respectively. Enlarged figures and additional example qualitative results for the ADNI dataset are available in the Supplementary. In Figure \ref{fig:example_UKBB_results} we see that whilst all models are able to detect the lesion visible in the sagittal slice, the VAE, cVAE, LDM and LDM ($T_{avg}$) produce very smooth outputs or lose defining characteristics and thus exhibit more false positives in healthy tissue. THOR and CADD provide the best results, with CADD better detecting white matter hypointensities (WMH). However, neither method fully inpaints all WMH, potentially due to presence of WMH in the healthy training set.


For the UoTH dataset, Table \ref{tab:disease_detection_results} shows that whilst CADD outperforms all other DDPM methods, it is outperformed by the VAE and cVAE models. Unlike the UKBB and ADNI datasets, the UoTH dataset contains noisy images with larger lesions and regions of pathology. Here, the CADD threshold, which limits the number of regions flagged as anomalous at each inpainting step, may be too stringent to fully inpaint extensive anomalies. It should be noted, however, that the improved disease detection performance of the VAE and cVAE models comes at the cost of accurate reconstruction of healthy tissue as illustrated in Table \ref{tab:image_quality_results} and Figure \ref{fig:example_UoT_results}. Such poor quality reconstructions would not be suitable for downstream tasks such as anomaly segmentation or image processing algorithms. 

\noindent\textbf{Encoding disease-related effects.} Table \ref{tab:corr_results_ADNI} shows the $\rho$ values of z-score (MAE (1\%)) with composite cognitive scores. Our proposed model demonstrates competitive performance, achieving the best or second best $\rho$ with three of the four cognitive measures and best overall rank.


\begin{table}
    \centering
    \smaller 
    \setlength{\abovecaptionskip}{3pt} 
    \setlength{\belowcaptionskip}{0pt} 
    \begin{tabularx}{\columnwidth}{p{2cm}XXXXp{0.3cm}} 
        \hline
        Method & $\rho_{\text{M}}$ & $\rho_{\text{EF}}$ & $\rho_{\text{LAN}}$ & $\rho_{\text{VIS}}$ & Rank \\
        \hline
        VAE & -0.4039 & -0.4212 & -0.3243 & \textbf{-0.2991} & 3.50 \\
        cVAE & \underline{-0.4311} & \underline{-0.4384} & -0.3342 & -0.2816 & 2.50 \\
        LDM & -0.3861 & -0.3944 & -0.3036 & -0.2169 & 6.50 \\
        LDM ($T_{\text{avg}}$) & -0.4029 & -0.4074 & -0.2976 & -0.2251 & 5.75 \\
        AutoDDPM (3D) & -0.4008 & -0.3746 & -0.2790 & -0.2096 & 5.50 \\ 
        THOR (3D) & \textbf{-0.4360} & -0.4260 & \textbf{-0.3436} & -0.2704 & \underline{2.25} \\
        CADD (Ours) & -0.4306 & \textbf{-0.4558} & \underline{-0.3376} & \underline{-0.2856} & \textbf{2.00} \\
        \hline
    \end{tabularx}
    \caption{Detecting disease effects in the ADNI dataset using CADD and baseline methods. $\rho_{\text{M}}$, $\rho_{\text{EF}}$, $\rho_{\text{LAN}}$, and $\rho_{\text{VS}}$ are the $\rho$ ($\downarrow$) values for the memory, executive function, language, and visuospatial functioning scores respectively.}
    \label{tab:corr_results_ADNI}
\end{table}

\begin{table}
    \centering
    \smaller 
    \setlength{\abovecaptionskip}{3pt} 
    \setlength{\belowcaptionskip}{0pt} 
    \begin{tabularx}{\columnwidth}{p{0.1cm}p{0.1cm}p{0.1cm}p{0.92cm}p{0.92cm}p{0.92cm}XXX}
        \hline
        & & & \multicolumn{3}{c}{MAE*LPIPS ($\downarrow$)} & \multicolumn{3}{c}{AUC ($\uparrow$)} \\
        \hline
        (1) & (2) & (3) & UKBB & ADNI & UoTH & UKBB & ADNI & UoTH \\
        \hline
        \checkmark & &  & 2.80E-3 & 5.22E-3 & 9.75E-3 & 0.5620 & 0.5641 & 0.4975 \\
        \checkmark & \checkmark &  & \underline{4.26E-4} & \underline{6.94E-4} & \textbf{8.74E-4} & \underline{0.6051} & \textbf{0.5866} & \underline{0.7144} \\
        \checkmark &  & \checkmark  & 2.82E-3 & 5.15E-3 & 7.93E-3 & 0.5735 & 0.5647 & 0.6000 \\
        \checkmark & \checkmark & \checkmark & \textbf{4.25E-4} & \textbf{6.92E-4} & \underline{1.01E-3} & \textbf{0.6052} & \underline{0.5847} & \textbf{0.7631} \\
        \hline
    \end{tabularx}

    \vspace{0.1cm}

    \begin{tabularx}{\columnwidth}{p{0.1cm}p{0.1cm}p{0.3cm}XXXX}
        \hline
        (1) & (2) & (3) & $\rho_{\text{M}}\,(\downarrow)$ & $\rho_{\text{EF}}\,(\downarrow)$ & $\rho_{\text{LAN}}\,(\downarrow)$ & $\rho_{\text{VIS}}\,(\downarrow)$\\
        \hline
        \checkmark &  &  & -0.4031 & -0.4194 & -0.3235 & -0.2657 \\
        \checkmark & \checkmark & & \underline{-0.4301} & \underline{-0.4453} & \underline{-0.3321} & \underline{-0.2782} \\ 
        \checkmark &  & \checkmark & -0.4241 & -0.4255 & -0.3301 & -0.2672 \\
        \checkmark & \checkmark & \checkmark & \textbf{-0.4306} & \textbf{-0.4558} & \textbf{-0.3376} & \textbf{-0.2856} \\
        \hline
    \end{tabularx}
    \caption{Ablation study results. Top: image quality and disease detection, MAE*LPIPS and AUC values respectively. AUC values are calculated from the z-score MAE (1\%) metric. Bottom: detecting disease effects. (1) CAAD backbone, (2) including inpainting scheme, (3) including clinical covariate conditioning.}
    \label{tab:ablation_study}
\end{table}

\subsection{Ablation studies}
Our ablation studies (Table \ref{tab:ablation_study}) highlight the impact of inpainting and clinical covariate conditioning components in CADD. Incorporating the inpainting scheme improves the MAE*LPIPS and AUC across all datasets, underscoring its importance in diffusion model-based anomaly detection. For these metrics, incorporating contextual clinical information improves the model performance in most scenarios. Both components lead to improvements in $\rho$ across all cognitive measures suggesting that these elements improve CADDs ability to detect disease specific effects. 

To further distinguish the performance gains of the CADD inpainting scheme from the effects of clinical conditioning, we evaluate CADD, THOR (3D), and AutoDDPM (3D) on the BraTS tumor segmentation dataset \cite{Baid2021}, for which clinical covariates are not available. We use FLAIR images (due to low contrast between lesions and healthy tissue in the BraTS T1-weighted scans) and first finetune the AutoencoderKL and DDPM elements of our model on 134 FLAIR images (with limited white matter hyperintensities) from the AIBL dataset \cite{Fowler2021}. We use a validation cohort of 33 subjects for early stopping and calculating $\text{KL}_{P_{95}}^{\text{val}}$. CADD still achieves SOTA results (Table \ref{tab:BraTS results}), both without clinical covariates and in the previously untested scenario involving large, varied lesion sizes, highlighting the versatility of our method. Moreover, CADD outperforms AutoDDPM (3D) and THOR (3D) for the UKBB and ADNI datasets even when covariates are incorporated into the frameworks of baseline models (see Supp. Table \ref{tab:image_quality_and_disease_results}). 

\begin{table}[!h]
    \centering
    \smaller
    \setlength{\abovecaptionskip}{3pt} 
    \setlength{\belowcaptionskip}{0pt} 
    \begin{tabularx}{\columnwidth}{Xp{2cm}p{1.8cm}p{1.8cm}} 
        \hline
        Metric & AutoDDPM (3D) & THOR (3D) & CADD (Ours) \\
        \hline
        Dice ($\uparrow$) & \underline{0.3386} & 0.2619 & \textbf{0.3548} \\ 
        \hline
    \end{tabularx}
    \caption{Average maximum dice for the BraTS dataset between anomaly maps and binarised ground-truth segmentation maps.}
    \label{tab:BraTS results}
\end{table}

\subsection{Limitations and future work}
Currently, CADD uses a fixed threshold to determine healthy and unhealthy regions during the reconstruction phase. However, differences in anomaly size and severity may require more adaptable thresholds. Future work will focus on developing a flexible inpainting scheme for diverse anomalies. Furthermore, for optimal results on a clinical dataset, CADD requires the full noising chain at inference time which can be time intensive. Future work will explore advances in fast sampling algorithms for diffusion models. There are many other factors which could contain relevant contextual information. Additional variables such as genetics, environmental features, scanner or site information, will be incorporated in further work.

\section{Conclusion}

We introduced CADD, the first conditional diffusion model framework for normative modeling in 3D brain images. By integrating clinical context and a reconstruction inpainting scheme, CADD achieves state-of-the-art performance in detecting neurological anomalies while preserving healthy brain features. Unlike prior models focused on large or artificial lesions, CADD effectively identifies disease effects in common neurological diseases and proves applicable to clinical data, demonstrating strong potential for real-world use. CADD ranks highest in image quality, disease detection, and disease-specific encoding tasks producing high-quality pseudo-healthy images which could enhance diagnosis and early intervention, and are applicable to downstream tasks like anomaly segmentation and image analysis.

\section{Acknowledgements}
This work was primarily funded by the JSPS Postdoctoral Fellowships for Research in Japan. Additional support was provided by the EPSRC-funded UCL Centre for Doctoral Training in Intelligent, Integrated Imaging in Healthcare (i4health) and the Department of Health’s NIHR-funded Biomedical Research Centre at University College London Hospitals and the NIH grants 1RF1AG080371, 1RF1MH123195, 1R01AG070988, 1R01EB031114, 1UM1MH130981, and 1R21NS138995.
\bibliographystyle{abbrv}
\bibliography{main}

\clearpage
\setcounter{page}{1}
\maketitlesupplementary

\section{Data pre-processing}

Each 3D brain image was pre-processed using the following pre-processing steps. The ANTS package (\url{https://stnava.github.io/ANTs/}) was used for affine registration of the images to the MNI 152 brain template. Images were then resampled to 130 × 130 × 130 resolution. Simple ITK (\url{https://github.com/SimpleITK/SimpleITK}) was used to perform n4 bias field correction and HD-BET (\url{https://github.com/MIC-DKFZ/HD-BET}) was used to skull strip the images. Following pre-processing, each image was resized to 128 x 128 x 128 and the pixel values were normalized to be between 0 and 1.

\section{CADD Transformer backbone framework}

Here we provide a more detailed description of the CADD diffusion model Transformer backbone. Consider a brain image in the latent space $\mathbf{z} \in \mathbb{R}^{h \times w \times d \times c}$. We first translate $\mathbf{z}$ into a sequence of tokens denoted as $\hat{\mathbf{k}} \in \mathbb{R}^{n_h \times n_w \times n_d \times L}$ such that the there are a total of $n_h \times n_w \times n_d$ $L$-dimensional tokens. We incorporate an absolute positional encoding \cite{vaswani2023}, $\mathbf{p}$, such that the input for the Transformer backbone becomes $\mathbf{k}=\hat{\mathbf{k}}+\mathbf{p}$. For input into the first transformer block, $\mathbf{z}$ is reshaped into $\mathbf{z}_{hw}\in \mathbb{R}^{n_d \times d \times L}$  where $d = n_h \times _w$ denotes the token count for each depth index. The Transformer block output is subsequently reshaped into $\mathbf{z}_d  \in \mathbb{R}^{d \times n_d \times L}$ to serve as input for the second Transformer block. The first Transformer block is designed to capture spatial information at a specific depth within the latent space, while the second Transformer block captures spatial information across tokens extracted from different depth indices. To embed the first two spatial dimensions into tokens, we apply the patch embedding technique outlined in ViT \cite{Dosovitskiy2021} for $n_d$. 

\section{Comparison methods implementations}

In this section we provide further details regarding the implementation of baseline methods. 

\subsection{LDM and LDM ($T_{\text{avg}}$)} For the LDM and LDM ($T_{\text{avg}}$) methods, we use the code available at: \url{https://github.com/marksgraham/ddpm-ood} for training and inference. As in the original paper \cite{graham2023c}, we use the PLMS scheduler to generate reconstructions. 

\subsection{VAE and cVAE} For the VAE and cVAE methods, we extend the 2D architecture provided at: \url{https://github.com/StefanDenn3r/Unsupervised_Anomaly_Detection_Brain_MRI} to 3D by converting 2D convolutions in the encoder and decoder blocks to 3D. 

\subsection{AutoDDPM (3D)} In this work, we extend the original AutoDDPM \cite{Bercea2023b} implementation to 3D. To make the AutoDDPM framework computationally viable for 3D images, we build the DDPM in the latent space of the AutoencoderKL CADD first stage model. We use the same DDPM for AutoDDPM (3D) as we do for CADD. As such, we conduct the AutoDDPM masking, stitching and resampling procedures applied at inference time, in the latent rather than image space. As in the original paper, we noise to an intermediary noise level of  $T_{\text{int}}=200$. The original AutoDDPM process generates the following mask between original $\mathbf{x}$ and reconstructed $\hat{\mathbf{x}}_0$ images:
\begin{equation}
   \hat{m}=\operatorname{norm}_{95}\left(\left|\hat{\mathbf{x}}_0-\mathbf{x}\right|\right) * \mathcal{S}_{l p i p s}\left(\hat{\mathbf{x}}_0, \mathbf{x}\right)
\end{equation}
where $\hat{m}$ is a binary mask. Here, we instead generate the following mask in the latent space:
\begin{equation}
   \hat{m}=\operatorname{norm}_{95}\left(\left|\hat{\mathbf{z}}_0-\mathbf{z}\right|\right)
\end{equation}
omitting the $\mathcal{S}_{l p i p s}$ similarity metric as this is a image-space metric. Following \cite{Bercea2023b}, the mask is applied for a stitching and re-sampling process with $T=50$. We conduct 4 re-sampling steps with seeds 42, 12, 1, 90. We use the original code: \url{https://github.com/ci-ber/autoDDPM} to guide our implementation.

\subsection{THOR (3D)}
As with AutoDDPM (3D), to extend THOR to 3D images we use a LDM with the AutoencoderKL first-stage model and Transformer DDPM backbone. Unlike AutoDDPM (3D), as we do not require partial denoising to be carried out after the stitching process and so we conduct the masking and stitching in the image-space. We calculate DDPM reconstructions for noise levels at 50 step intervals up to $T_{\text{int}}=350$, as done in the original work \cite{Bercea2024c}. We calculate the following mask between the AutoencoderKL decoder output $\hat{\mathbf{x}}$ and DDPM reconstruction $\hat{\mathbf{x}}_0$: 
\begin{equation}
\hat{m}=\operatorname{norm}_{95}\left(\left|\hat{\mathbf{x}}_0-\hat{\mathbf{x}}\right|\right) * \mathcal{S}_{l p i p s}\left(\hat{\mathbf{x}}_0, \hat{\mathbf{x}}\right)
\end{equation}
which is then normalized between 0 and 1. A sequence of closing and dilation operations (termed $cd$) are then applied to the masks. We use the anomaly masks to mask healthy/unhealthy tissue:
\begin{equation}
    \hat{\mathbf{x}}^t =c d\left(m\left(\hat{\mathbf{x}}^t_0, \hat{\mathbf{x}}\right)\right) \cdot \hat{\mathbf{x}}^t_0+\left(1-c d\left(m\left(x_0^t, \hat{\mathbf{x}}\right)\right)\right) \cdot \hat{\mathbf{x}}.
\end{equation}
We average reconstructions from each $T$ value. We use the original code: \url{https://github.com/ci-ber/THOR_DDPM} to guide our implementation.

\section{Efficiency analysis} Table \ref{tab:efficiency results} shows that, of the inpainting methods, CADD has the second fastest sampling time and is considerably faster than THOR. It should be noted that the sampling time for all methods is reasonable when considering the amount of time required to generate a typical MR scan.
\begin{table}[!h]
    \centering
    \smaller
    \begin{tabularx}{\columnwidth}{Xp{2cm}p{1.6cm}p{1.6cm}} 
        \hline
        Metric & AutoDDPM (3D) & THOR (3D) & CADD (Ours) \\
        \hline
        Sampling time (s) & 2.2901 & 8.1916 & 3.6883 \\
        \hline
    \end{tabularx}
    \caption{Sampling time for a single sample from the ADNI dataset.}
    \label{tab:efficiency results}
\end{table}

\section{Inclusion of clinical information}

In addition to the ablation studies in the main paper, we further illustrate that the improved performance of CADD compared to baselines is not solely due to the inclusion of clinical information in the modeling framework by repeating the experiments in Tables \ref{tab:image_quality_results} and \ref{tab:disease_detection_results} of the main paper for the UKBB and ADNI datasets incorporating covariates into THOR (3D) and AutoDDPM (3D). To do this, we follow the S-AdaLN approach used in CADD. Table \ref{tab:image_quality_and_disease_results} shows that CADD still outperforms both baselines even with the inclusion of clinical covariates. 
\begin{table*}[!tb]
    \centering
    \smaller
    \begin{tabularx}{\textwidth}{p{0.5cm}p{2.4cm}XXXp{2cm}p{2cm}p{2.1cm}}
        \hline
        Dataset & Method & MAE ($\downarrow$) & PSNR ($\uparrow$) & $\operatorname{SSIM}$ ($\uparrow$) & $\text{LPIPS}_{\text{alex}}$ ($\downarrow$) & $\text{LPIPS}_{\text{vgg}}$ ($\downarrow$) & $\text{LPIPS}_{\text{squeeze}}$ ($\downarrow$) \\
        \hline
        \multirow{3}{*}{UKBB} 
        & AutoDDPM (3D) & 0.0229$\pm$0.0003 & 25.1185$\pm$0.1173 & 0.8646$\pm$0.0007 & 0.0797$\pm$0.0004 & 0.1235$\pm$0.0005 & 0.0662$\pm$0.0004 \\
        & THOR (3D) & 0.0114$\pm$0.0002 & 31.9226$\pm$0.1774 & 0.9503$\pm$0.0007 & 0.0584$\pm$0.0008 & 0.0837$\pm$0.0006 & 0.0524$\pm$0.0005 \\
        & CADD (Ours) & \textbf{0.0103$\pm$0.0001} & \textbf{32.1909$\pm$0.1206} & \textbf{0.9543$\pm$0.0003} & \textbf{0.0406$\pm$0.0003} & \textbf{0.0740$\pm$0.0003} & \textbf{0.0404$\pm$0.0003} \\
        \hline
        \multirow{3}{*}{ADNI} 
        & AutoDDPM (3D) & 0.0373$\pm$0.0010 & 20.7969$\pm$0.2593 & 0.8310$\pm$0.0034 & 0.0926$\pm$0.0022 & 0.1441$\pm$0.0024 & 0.0731$\pm$0.0019 \\  
        & THOR (3D) & 0.0223$\pm$0.0010 & 25.7974$\pm$0.4998 & 0.9164$\pm$0.0023 & 0.0858$\pm$0.0030 & 0.0999$\pm$0.0020 & 0.0669$\pm$0.0019 \\
        & CADD (Ours) & \textbf{0.0162$\pm$0.0006} & \textbf{28.0765$\pm$0.3532} & \textbf{0.9486$\pm$0.0027} & \textbf{0.0425$\pm$0.0016} & \textbf{0.0797$\pm$0.0025} & \textbf{0.0414$\pm$0.0017} \\
    \end{tabularx}
    \begin{tabularx}{\textwidth}{lp{2.4cm}XXXXXXXX}
        \hline
        \multicolumn{2}{c}{z-score type:} & \multicolumn{2}{c}{MAE (1\%)} & \multicolumn{2}{c}{MAE (5\%)} & \multicolumn{2}{c}{MAE*LPIPS (1\%)} & \multicolumn{2}{c}{MAE*LPIPS (5\%)} \\
        Dataset & Method & AUC & p-value & AUC & p-value & AUC & p-value & AUC & p-value \\
        \hline
        \multirow{3}{*}{UKBB} 
        & AutoDDPM (3D) & 0.5789 & 1.753E-05 & 0.5705 & 1.318E-04 & 0.5652 & 4.015E-04 & 0.5582 & 1.598E-03 \\ 
        & THOR (3D) & 0.5959 & 3.807E-07 & 0.5818 & 1.157E-05 & 0.5811 & 4.581E-05 & 0.5716 & 4.930E-04 \\ 
        & CADD (Ours) & \textbf{0.6052} & \textbf{6.828E-09} & \textbf{0.6017} & \textbf{2.586E-08} & \textbf{0.5994} & \textbf{5.585E-08} & \textbf{0.5967} & \textbf{1.771E-07} \\
        \hline
        \multirow{3}{*}{ADNI}  
        & AutoDDPM (3D) & 0.5566 & 1.2684E-1 & 0.5453 & 2.3844E-1 & 0.5745 & 6.2304E-2 & 0.5630 & 1.2483E-1 \\
        & THOR (3D) & 0.5617 & 1.4641E-2 & 0.5693 & 2.3017E-2 & 0.5901 & 2.8806E-3 & 0.5951 & 3.3229E-3 \\ 
        & CADD (Ours) & \textbf{0.5847} & \textbf{2.0000E-3} & \textbf{0.5962} & \textbf{3.0000E-3} & \textbf{0.6412} & \textbf{8.0570E-5} & \textbf{0.6408} & \textbf{1.7720E-4} \\ 
        \hline
    \end{tabularx}
    \caption{Image quality and disease detection results for CADD, THOR (3D) and AutoDDPM (3D) baselines incorporating clinical covariates.}
    \label{tab:image_quality_and_disease_results}
\end{table*}

\section{Timestep analysis}\label{sec:Timestep_analysis}
\begin{figure}
     \centering
     \includegraphics[width=\linewidth]{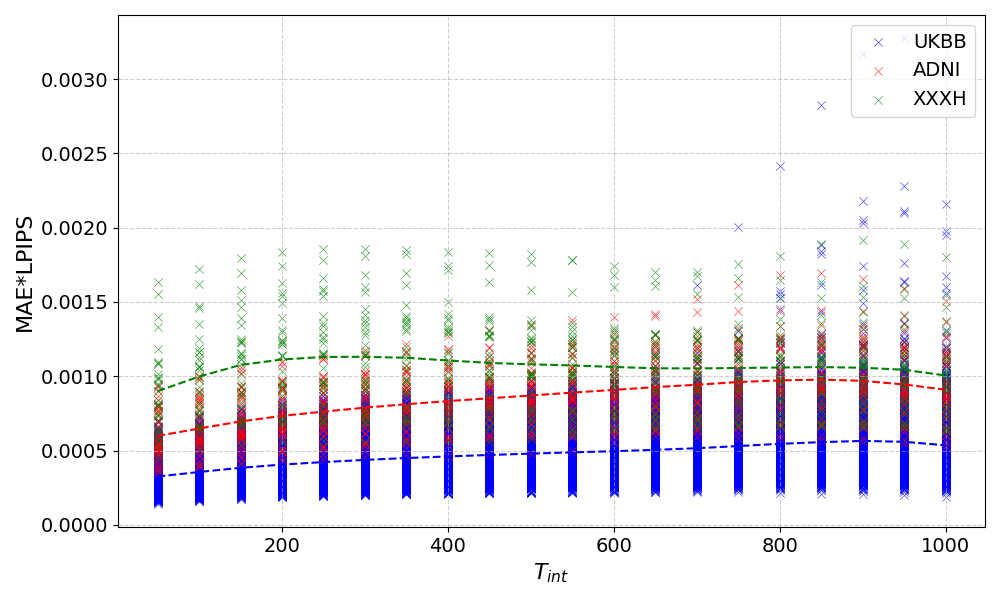}
     \caption{CADD MAE*LPIPS values for the healthy holdout dataset at different $T_{int}$ values.}
     \label{fig:mae_lpips}
 \end{figure}

\begin{figure}
     \centering
     \includegraphics[width=\linewidth]{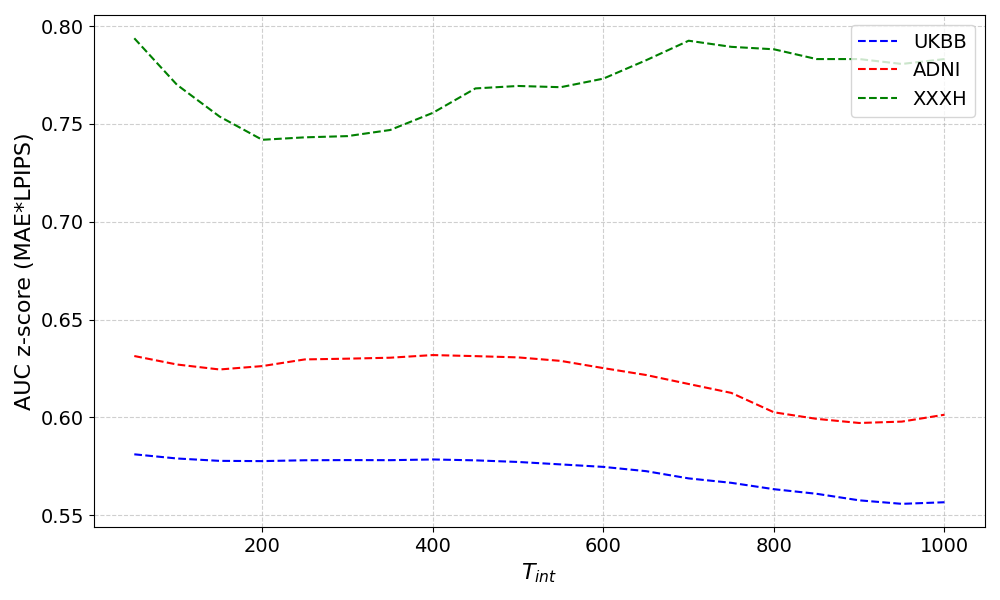}
     \caption{CADD AUC values between the healthy holdout cohort and disease cohort at different $T_{int}$ values.}
     \label{fig:auc}
 \end{figure}

 Figure \ref{fig:mae_lpips} provides the CADD MAE*LPIPS scores for the healthy cohort of each dataset for $T_{int} \in \{50 \cdot k : k = 1,\dots, \frac{1000}{50}\}$. Figure \ref{fig:auc} provides the CADD AUC scores between disease and healthy cohorts of each dataset using the average MAE*LPIPS values, z-scored using the holdout validation cohort, for $T_{int} \in \{50 \cdot k : k = 1,\dots, \frac{1000}{50}\}$.
 
\section{Qualitative image reconstruction results}
\begin{figure*}[ht]
    \centering
    \begin{subfigure}{0.12\linewidth} 
        \centering
        \includegraphics[width=\linewidth, trim=5 5 480 30, clip]{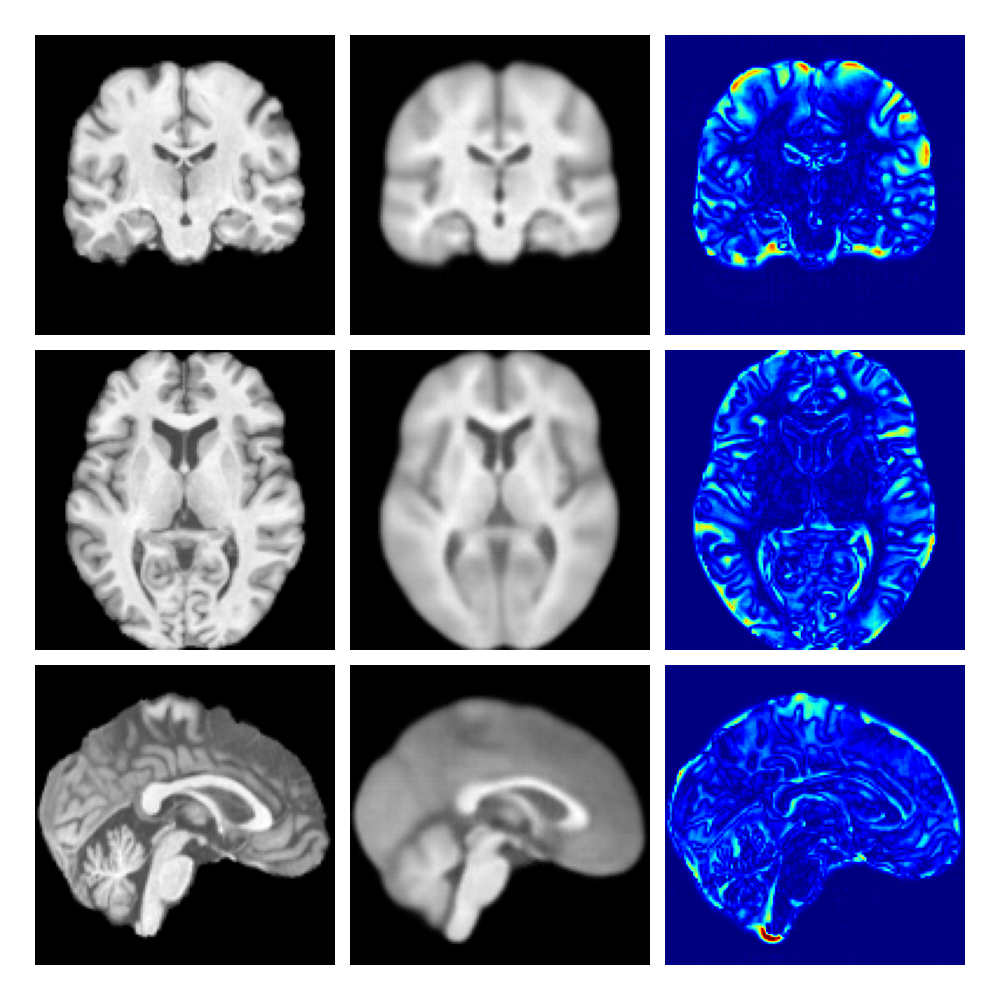}
        \caption{Input}
    \end{subfigure}
    \begin{subfigure}{0.24\linewidth} 
        \centering
        \includegraphics[width=\linewidth, trim=245 5 5 30, clip]{figures/ukbb/healthy/1033445_VAE_thresh06.png}
        \caption{VAE}
    \end{subfigure}
    \begin{subfigure}{0.24\linewidth}
        \centering
        \includegraphics[width=\linewidth, trim=245 5 5 30, clip]{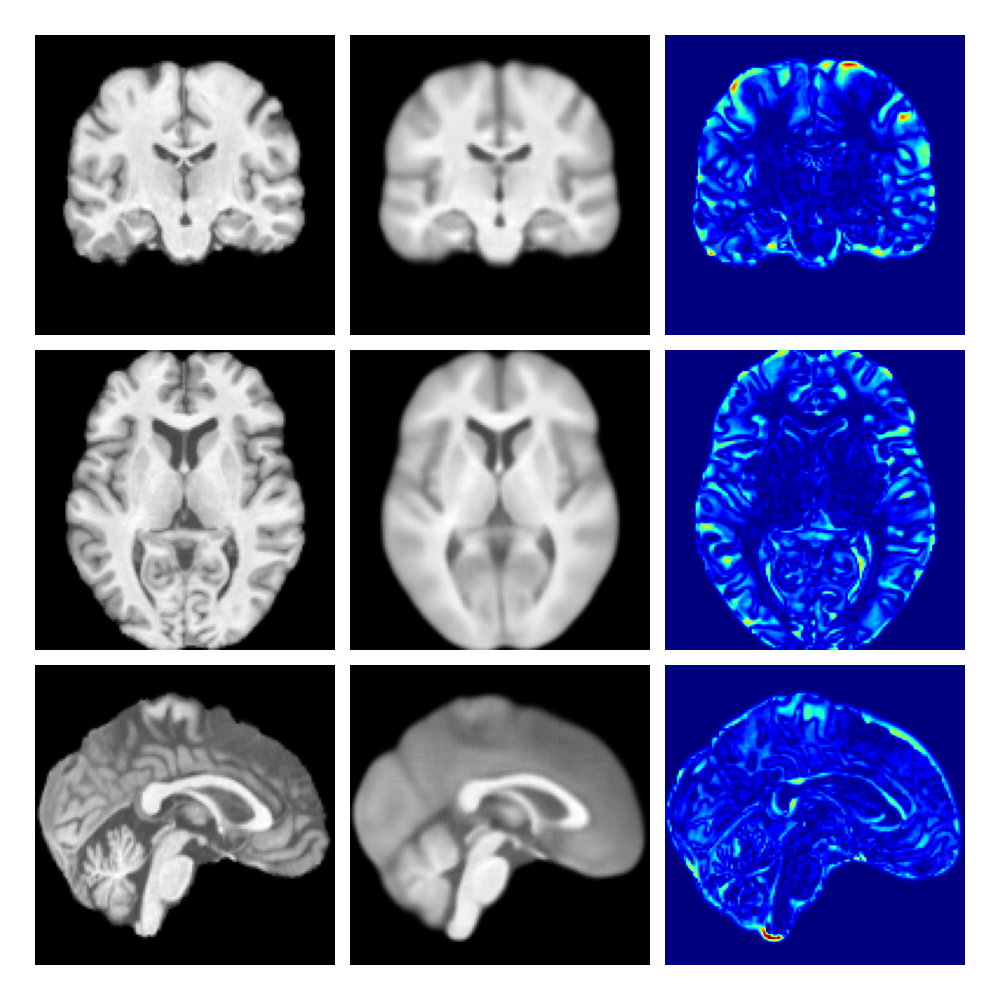}
        \caption{cVAE}
    \end{subfigure}
    \begin{subfigure}{0.24\linewidth}
        \centering
        \includegraphics[width=\linewidth, trim=245 5 5 30, clip]{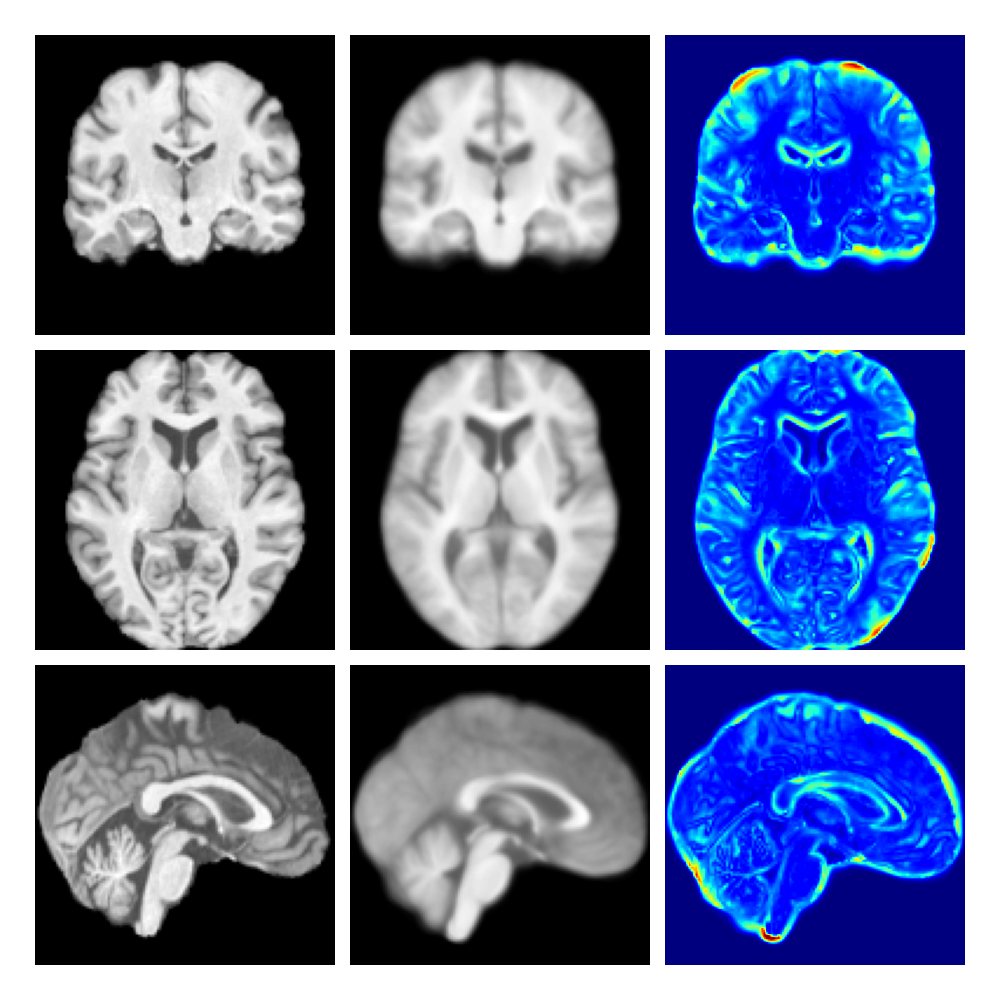}
        \caption{LDM ($T_{\text{avg}}$)}
    \end{subfigure}
    \begin{subfigure}{0.24\linewidth}
        \centering
        \includegraphics[width=\linewidth, trim=245 5 5 30, clip]{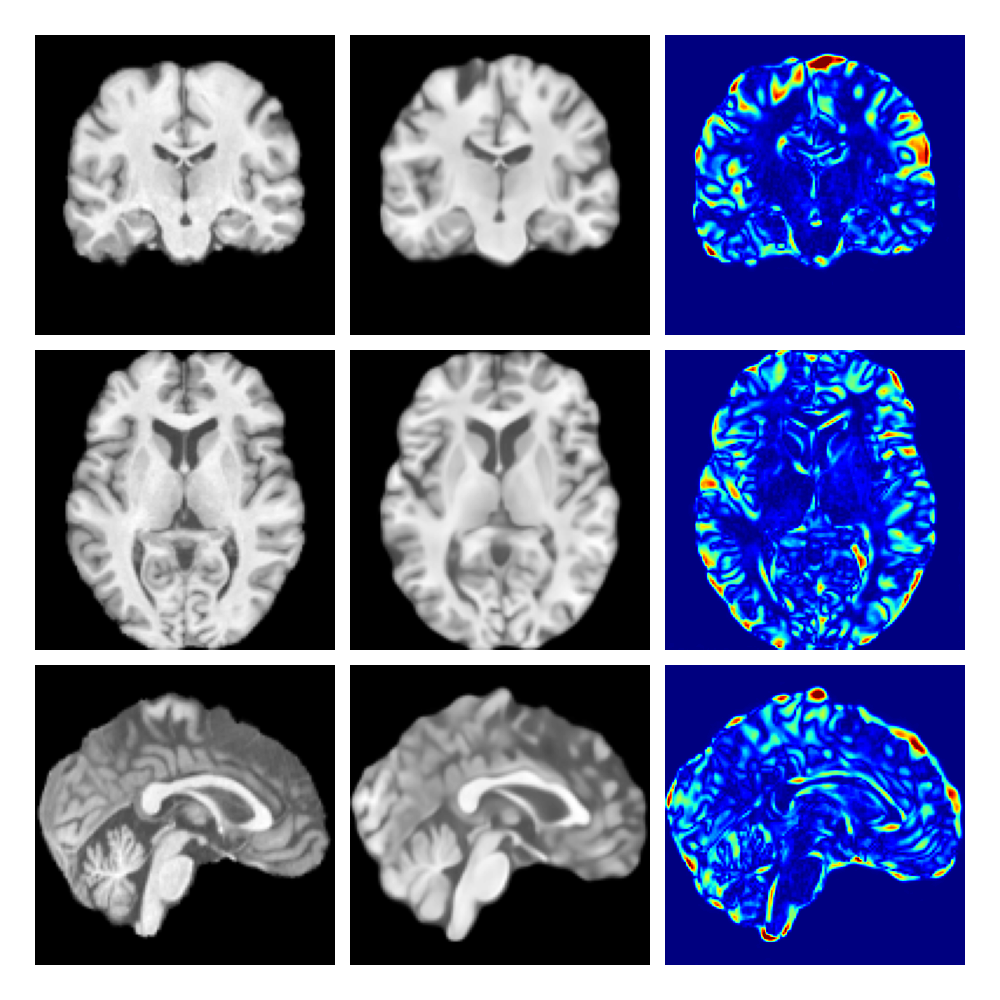}
        \caption{LDM}
    \end{subfigure}
    \begin{subfigure}{0.24\linewidth}
        \centering
        \includegraphics[width=\linewidth, trim=245 5 5 30, clip]{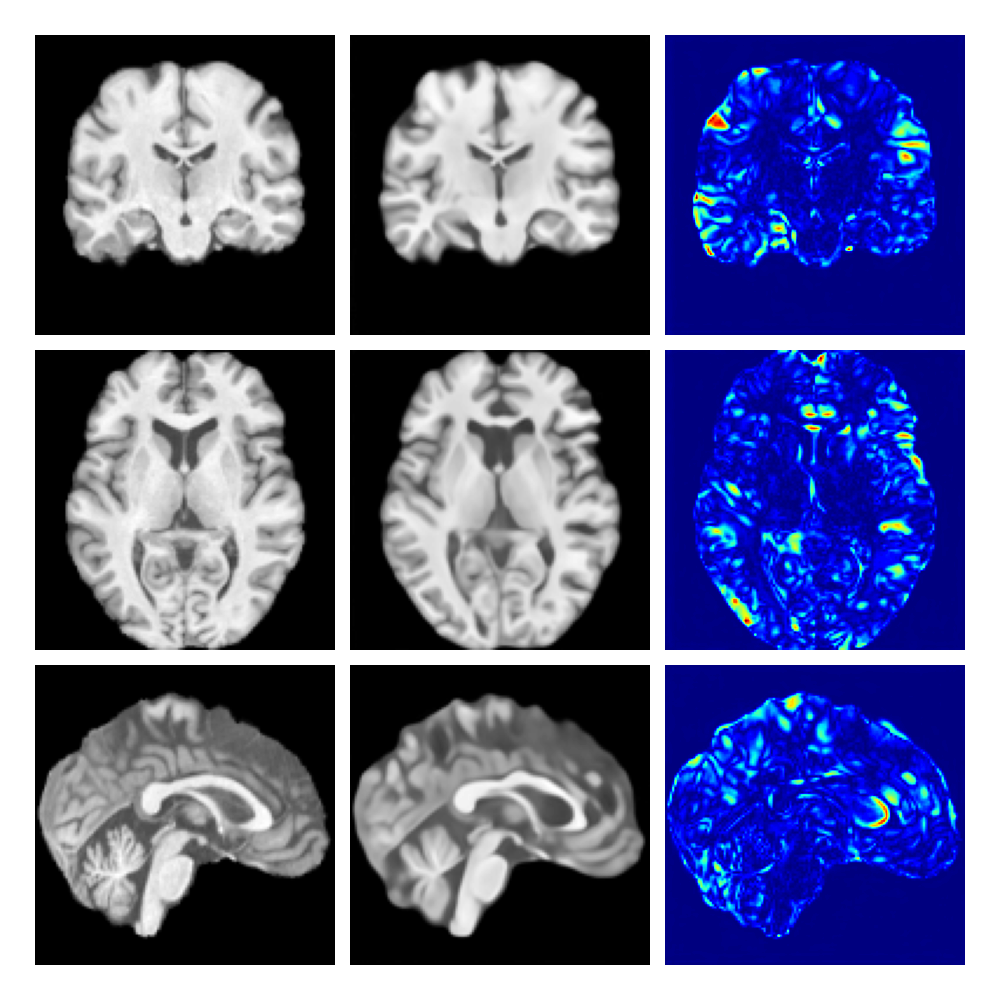}
        \caption{AutoDDPM}
    \end{subfigure}
    \begin{subfigure}{0.24\linewidth}
        \centering
        \includegraphics[width=\linewidth, trim=245 5 5 30, clip]{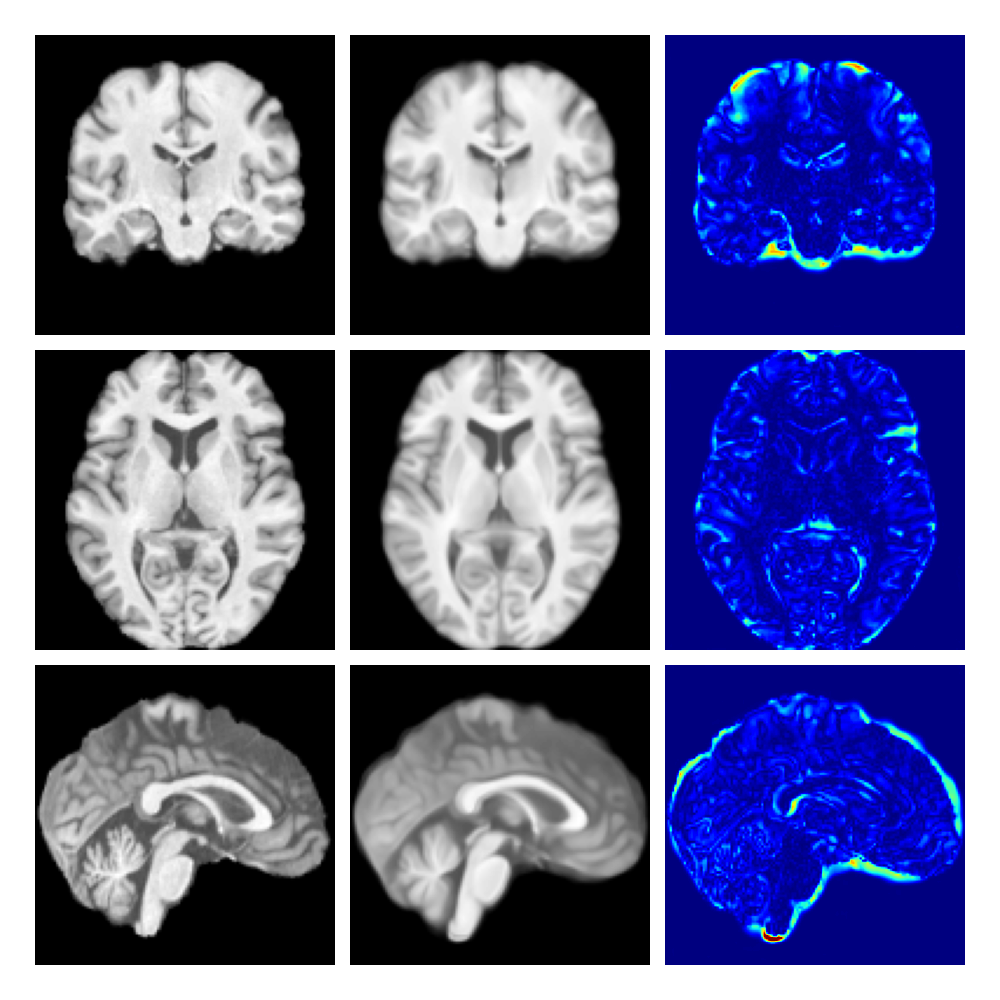}
        \caption{THOR}
    \end{subfigure}
    \begin{subfigure}{0.24\linewidth}
        \centering
        \includegraphics[width=\linewidth, trim=245 5 5 30, clip]{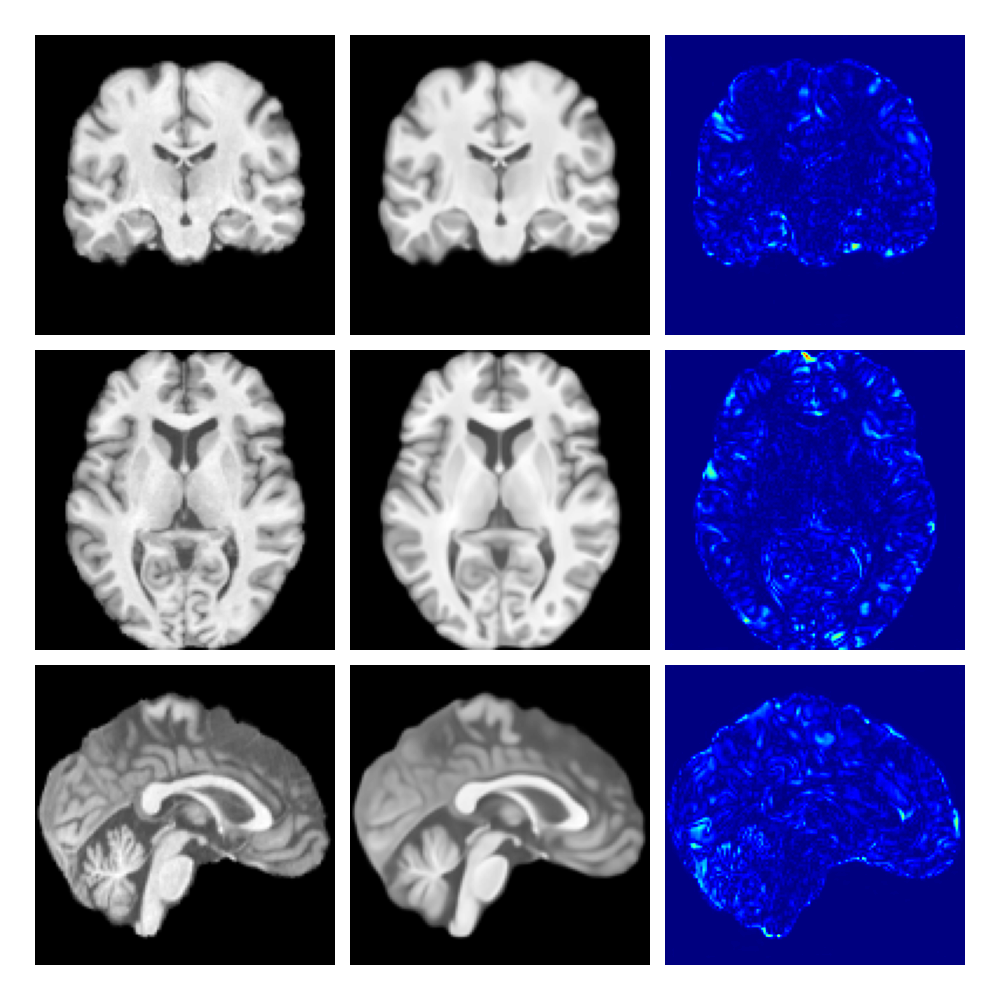}
        \caption{CADD (Ours)}
    \end{subfigure}
    \caption{Example healthy reconstructions and anomaly maps for a sample from the UK Biobank healthy test cohort. For a healthy subject, we should observe no regions highlighted in the anomaly map.}
    \label{fig:example_UKBB_healthy}
\end{figure*}

Figure \ref{fig:example_UKBB_healthy} provides example reconstructions and anomaly maps for a healthy subject from the UK Biobank holdout test cohort.

\section{Qualitative disease detection results}
\begin{figure*}[hbt!]
    \centering
    \begin{subfigure}{0.12\linewidth} 
        \centering
        \includegraphics[width=\linewidth, trim=5 5 666 30, clip]{figures/XXXH/disease/XF2F8071B19061DB9_VAE_thresh05.png}
        \caption{Input}
    \end{subfigure}
    \begin{subfigure}{0.24\linewidth} 
        \centering
        \includegraphics[width=\linewidth, trim=340 5 5 30, clip]{figures/XXXH/disease/XF2F8071B19061DB9_VAE_thresh05.png}
        \caption{VAE}
    \end{subfigure}
    \begin{subfigure}{0.24\linewidth}
        \centering
        \includegraphics[width=\linewidth, trim=245 5 5 30, clip]{figures/XXXH/disease/XF2F8071B19061DB9_cVAE_thresh05.png}
        \caption{cVAE}
    \end{subfigure}
    \begin{subfigure}{0.24\linewidth}
        \centering
        \includegraphics[width=\linewidth, trim=245 5 5 30, clip]{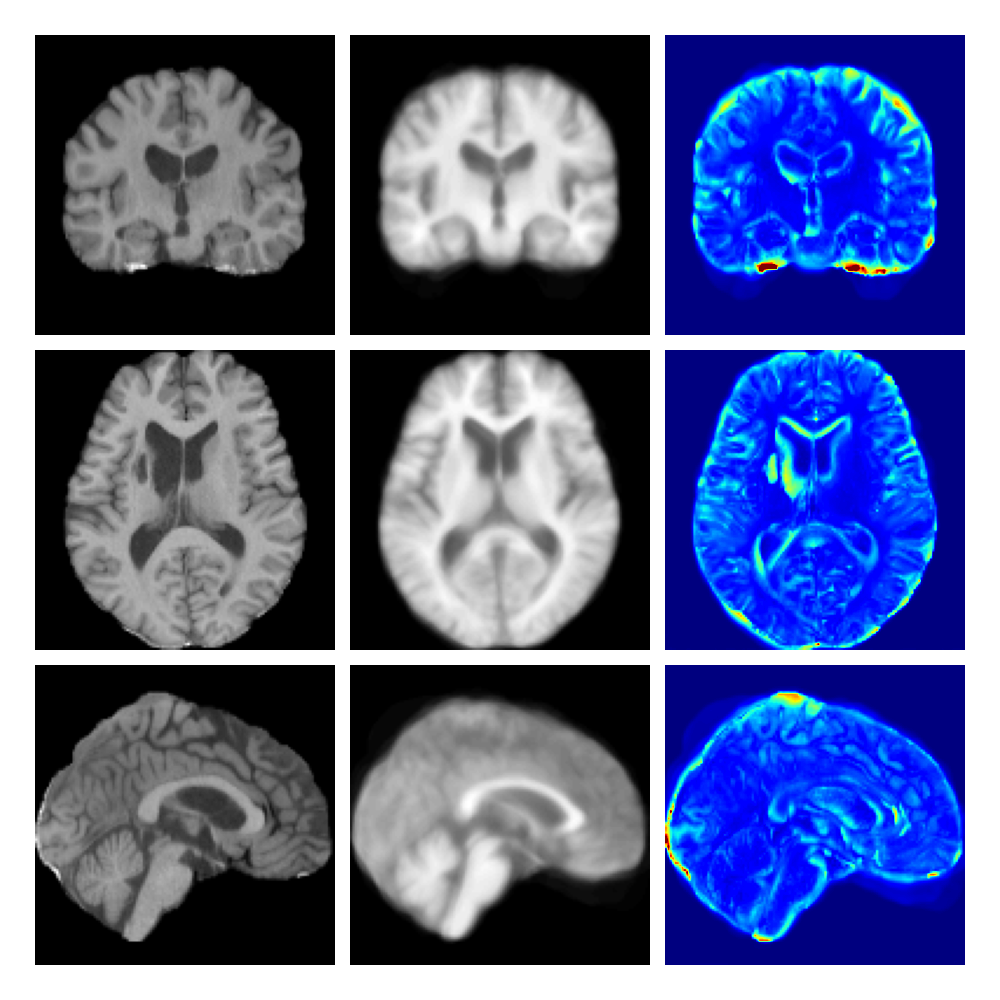}
        \caption{LDM ($T_{\text{avg}}$)}
    \end{subfigure}
    \begin{subfigure}{0.24\linewidth}
        \centering
        \includegraphics[width=\linewidth, trim=245 5 5 30, clip]{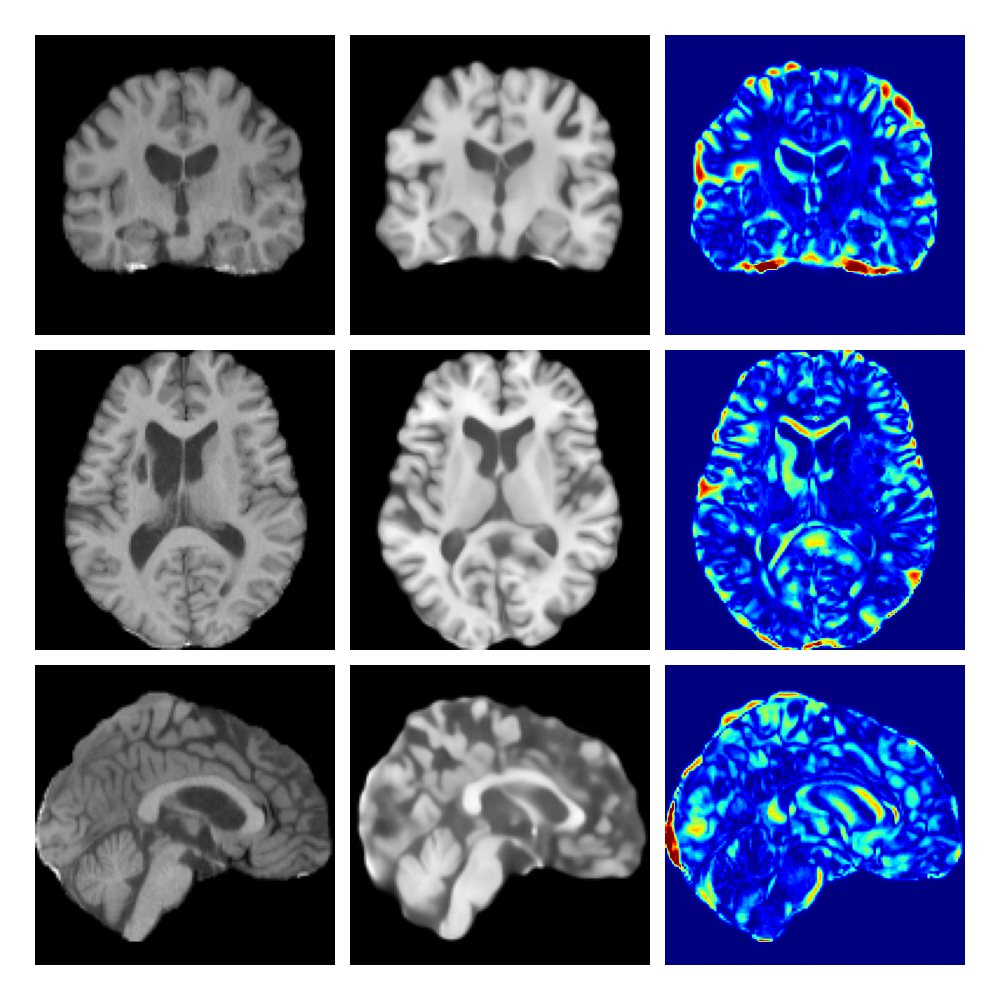}
        \caption{LDM}
    \end{subfigure}
    \begin{subfigure}{0.24\linewidth}
        \centering
        \includegraphics[width=\linewidth, trim=245 5 5 30, clip]{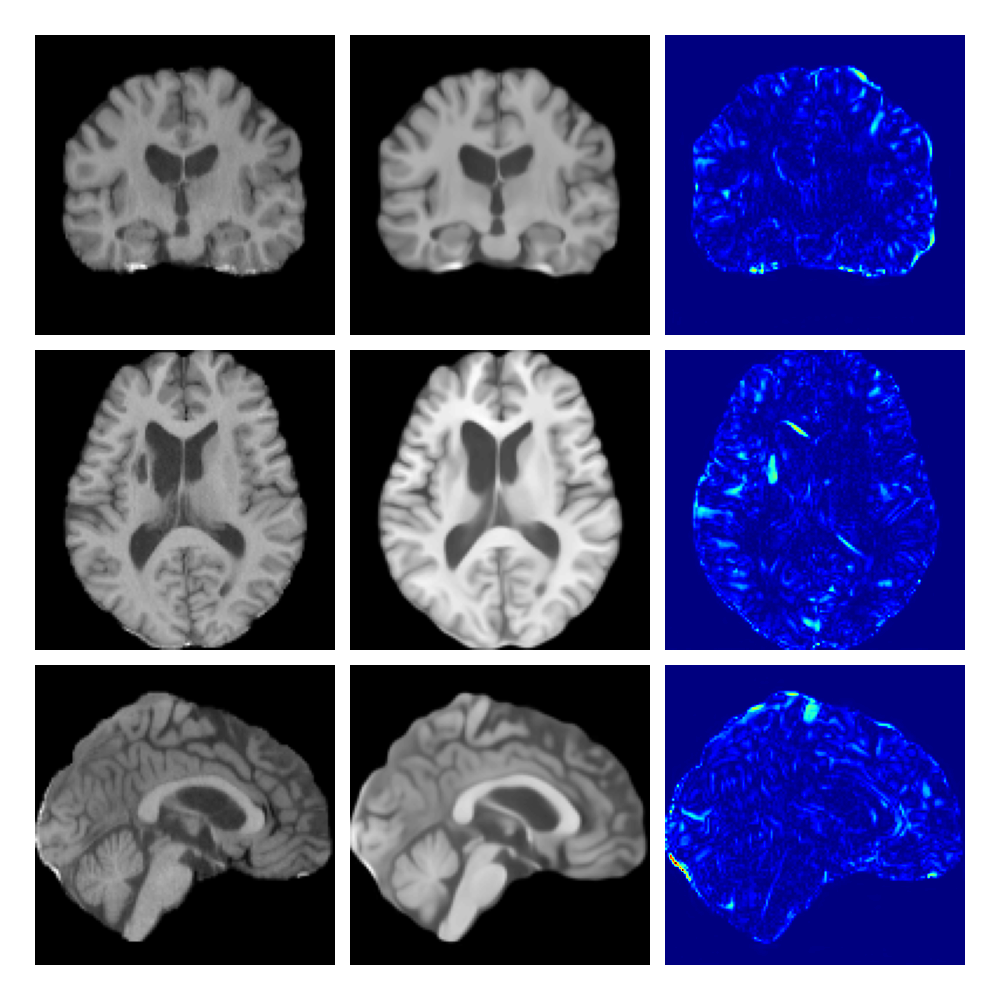}
        \caption{AutoDDPM}
    \end{subfigure}
    \begin{subfigure}{0.24\linewidth}
        \centering
        \includegraphics[width=\linewidth, trim=245 5 5 30, clip]{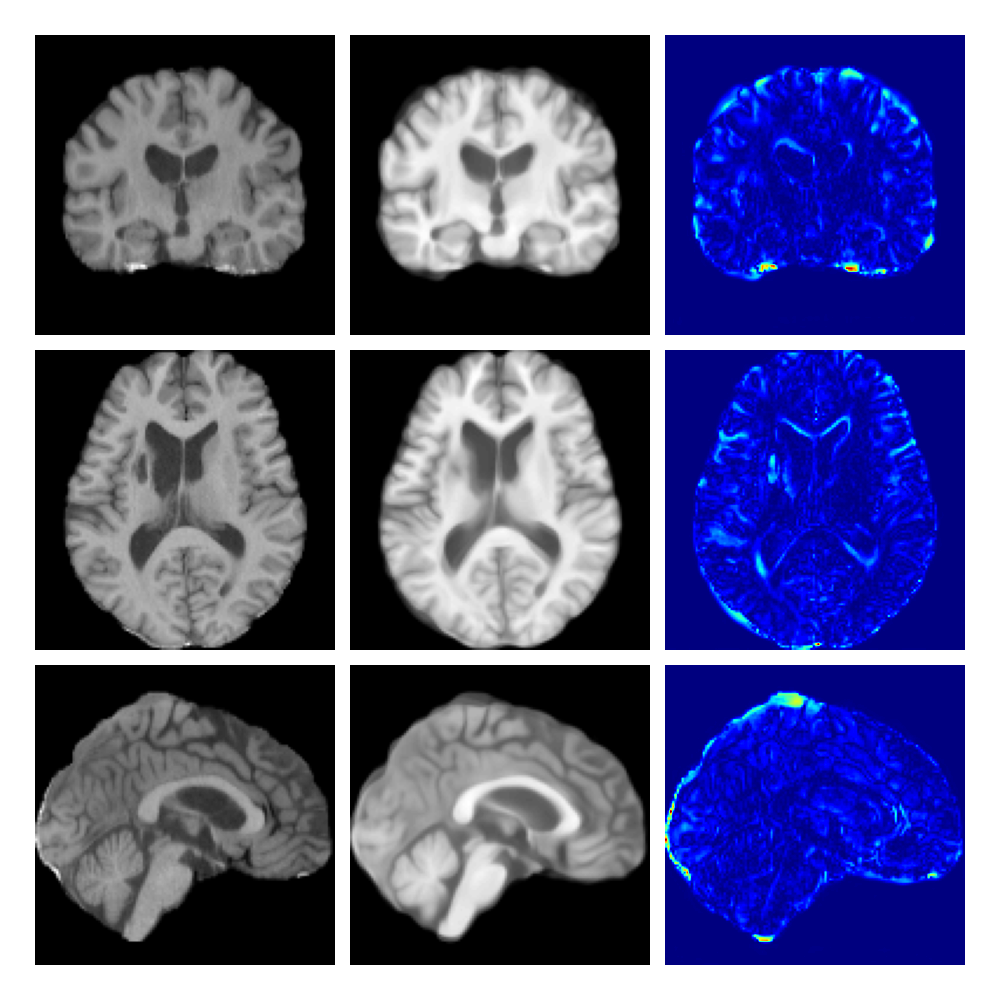}
        \caption{THOR}
    \end{subfigure}
    \begin{subfigure}{0.24\linewidth}
        \centering
        \includegraphics[width=\linewidth, trim=245 5 5 30, clip]{figures/XXXH/disease/XF2F8071B19061DB9_CADD_thresh05.png}
        \caption{CADD (Ours)}
    \end{subfigure}
    
    \caption{Enlarged example reconstructions and anomaly maps for a sample from the disease cohort of the XXXH dataset. The lesion region is indicated in the original image by the red box.}
    \label{fig:enlarged_example_UoT_results}
\end{figure*}

\begin{figure*}[hbt!]
    \centering
    \begin{subfigure}{0.12\linewidth} 
        \centering
        \includegraphics[width=\linewidth, trim=5 5 480 30, clip]{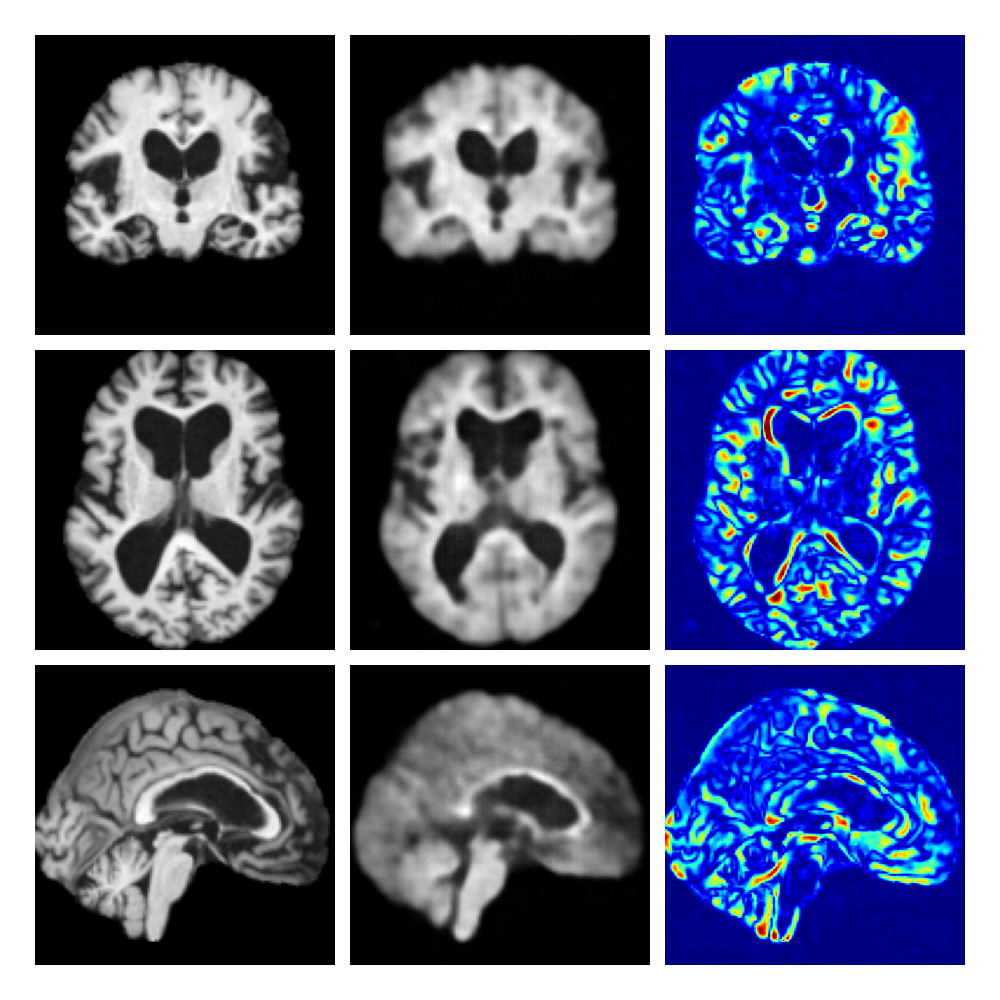}
        \caption{Input}
    \end{subfigure}
    \begin{subfigure}{0.24\linewidth} 
        \centering
        \includegraphics[width=\linewidth, trim=245 5 5 30, clip]{figures/ADNI/AD_1379_VAE_thresh06.png}
        \caption{VAE}
    \end{subfigure}
    \begin{subfigure}{0.24\linewidth}
        \centering
        \includegraphics[width=\linewidth, trim=245 5 5 30, clip]{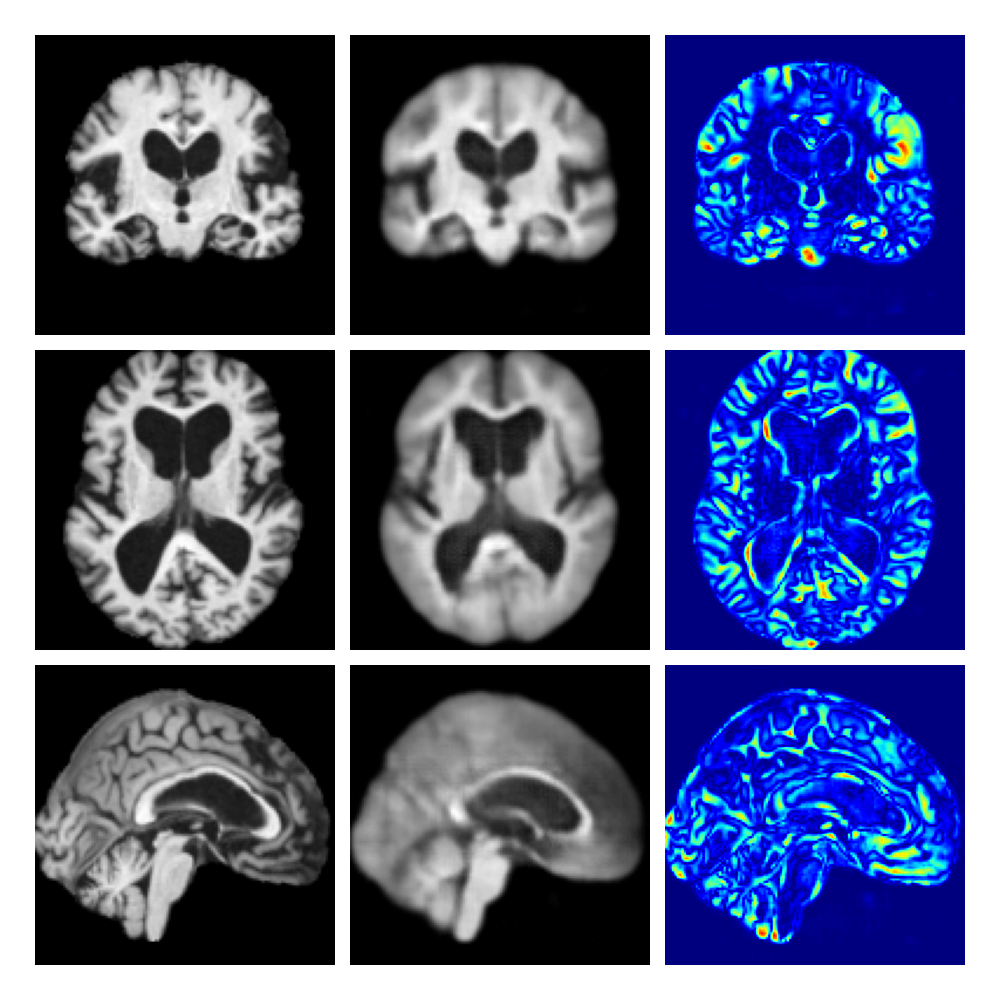}
        \caption{cVAE}
    \end{subfigure}
    \begin{subfigure}{0.24\linewidth}
        \centering
        \includegraphics[width=\linewidth, trim=245 5 5 30, clip]{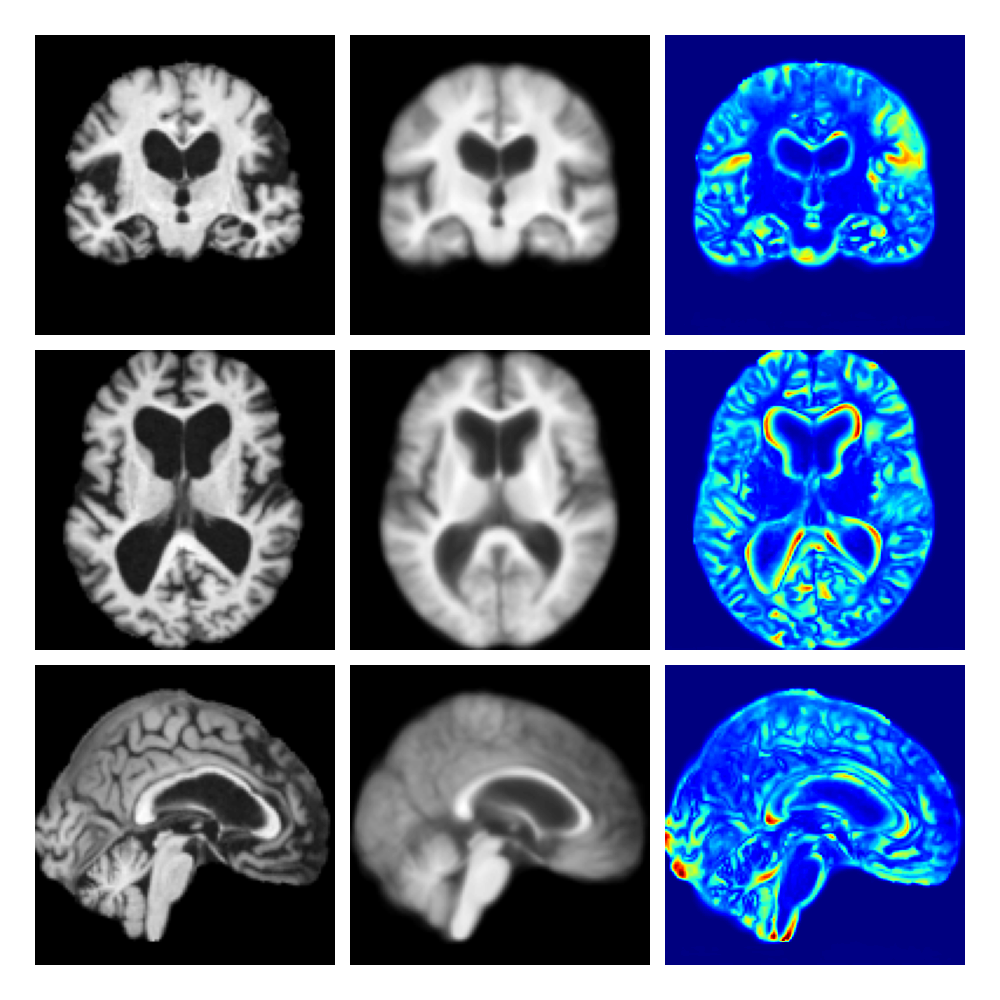}
        \caption{LDM ($T_{\text{avg}}$)}
    \end{subfigure}
    \begin{subfigure}{0.24\linewidth}
        \centering
        \includegraphics[width=\linewidth, trim=245 5 5 30, clip]{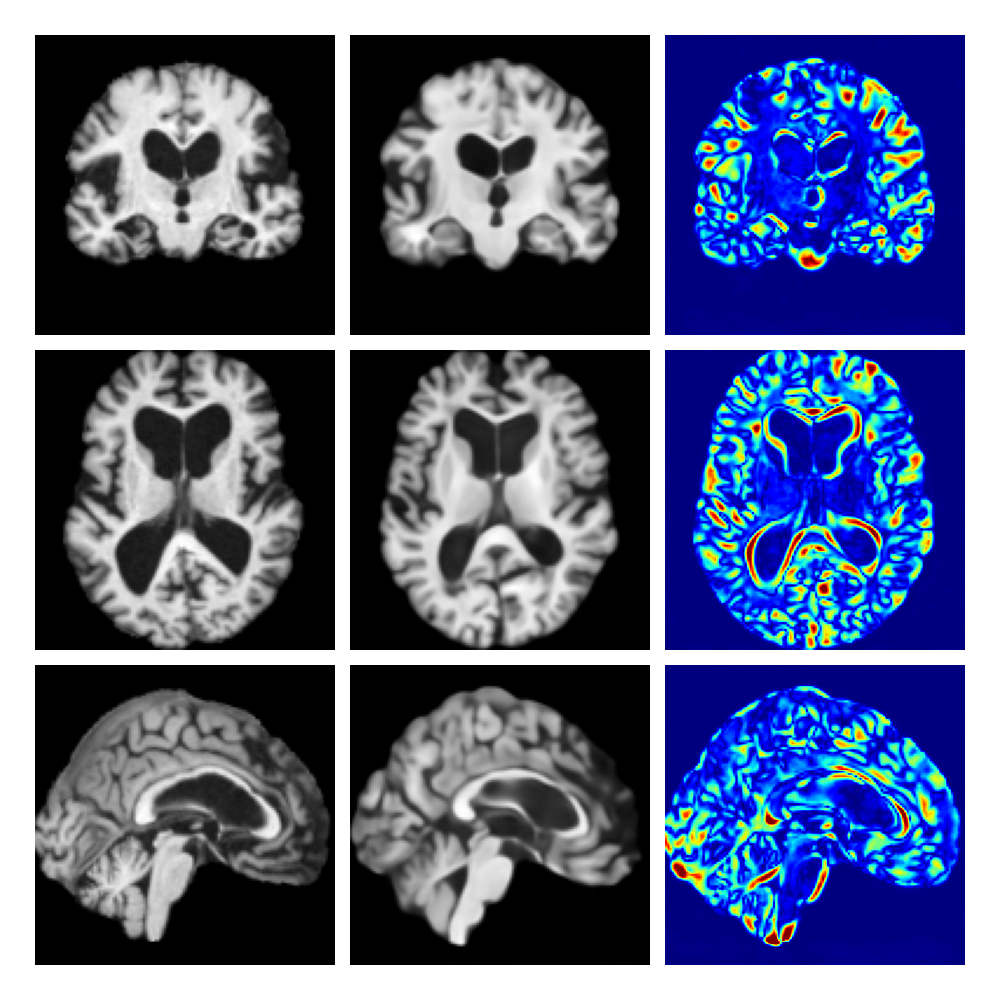}
        \caption{LDM}
    \end{subfigure}
    \begin{subfigure}{0.24\linewidth}
        \centering
        \includegraphics[width=\linewidth, trim=245 5 5 30, clip]{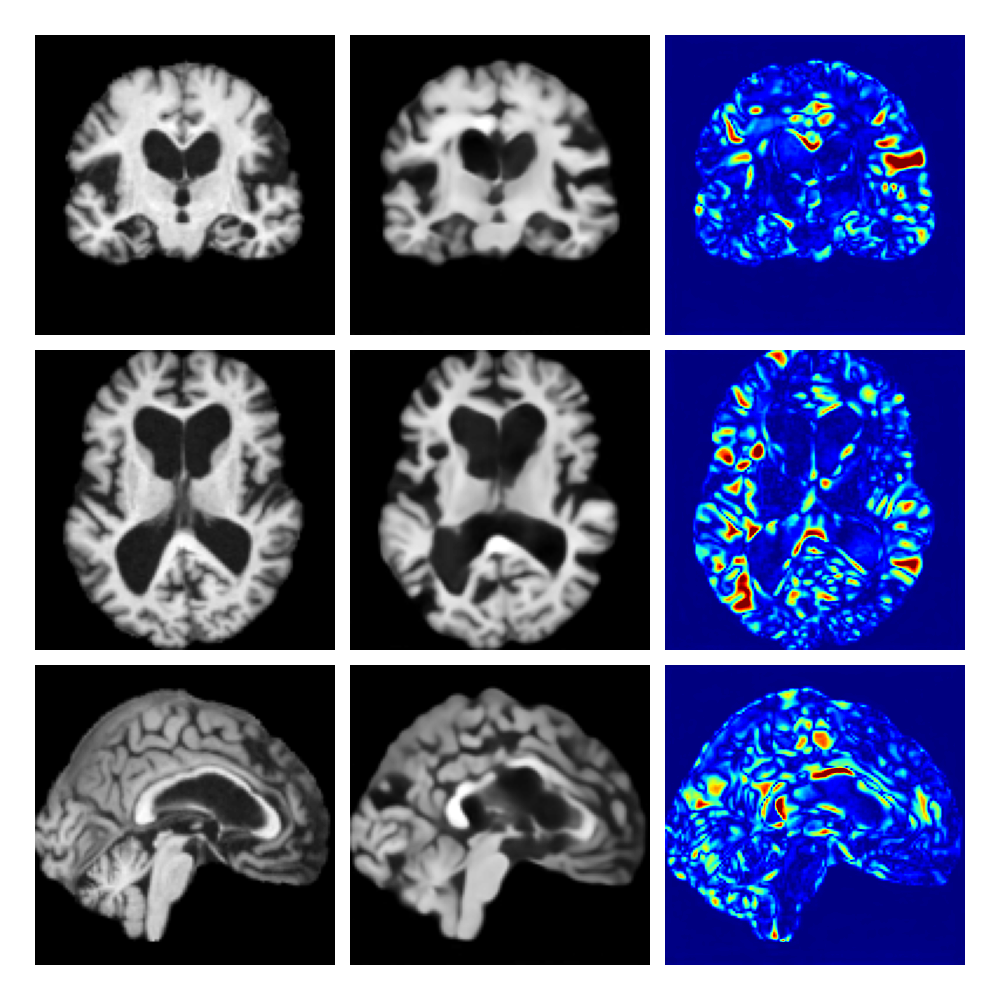}
        \caption{AutoDDPM}
    \end{subfigure}
    \begin{subfigure}{0.24\linewidth}
        \centering
        \includegraphics[width=\linewidth, trim=245 5 5 30, clip]{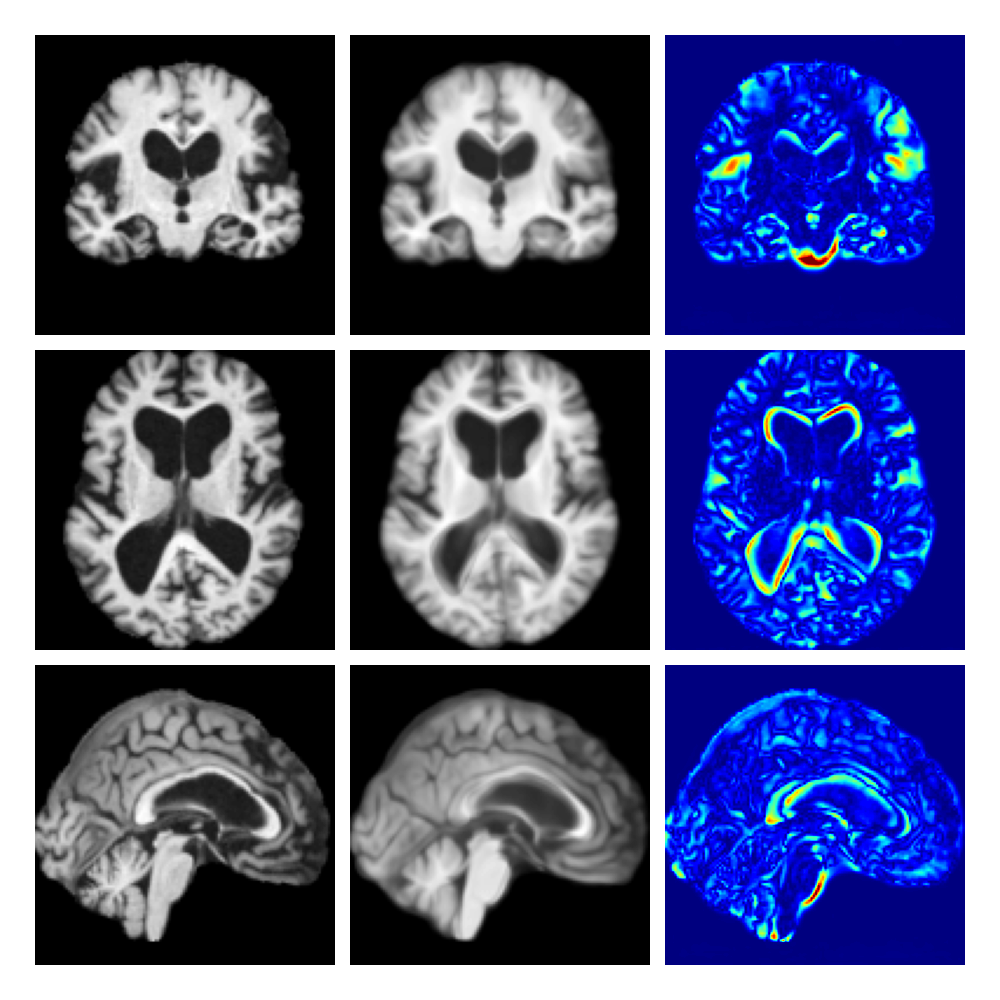}
        \caption{THOR}
    \end{subfigure}
    \begin{subfigure}{0.24\linewidth}
        \centering
        \includegraphics[width=\linewidth, trim=245 5 5 30, clip]{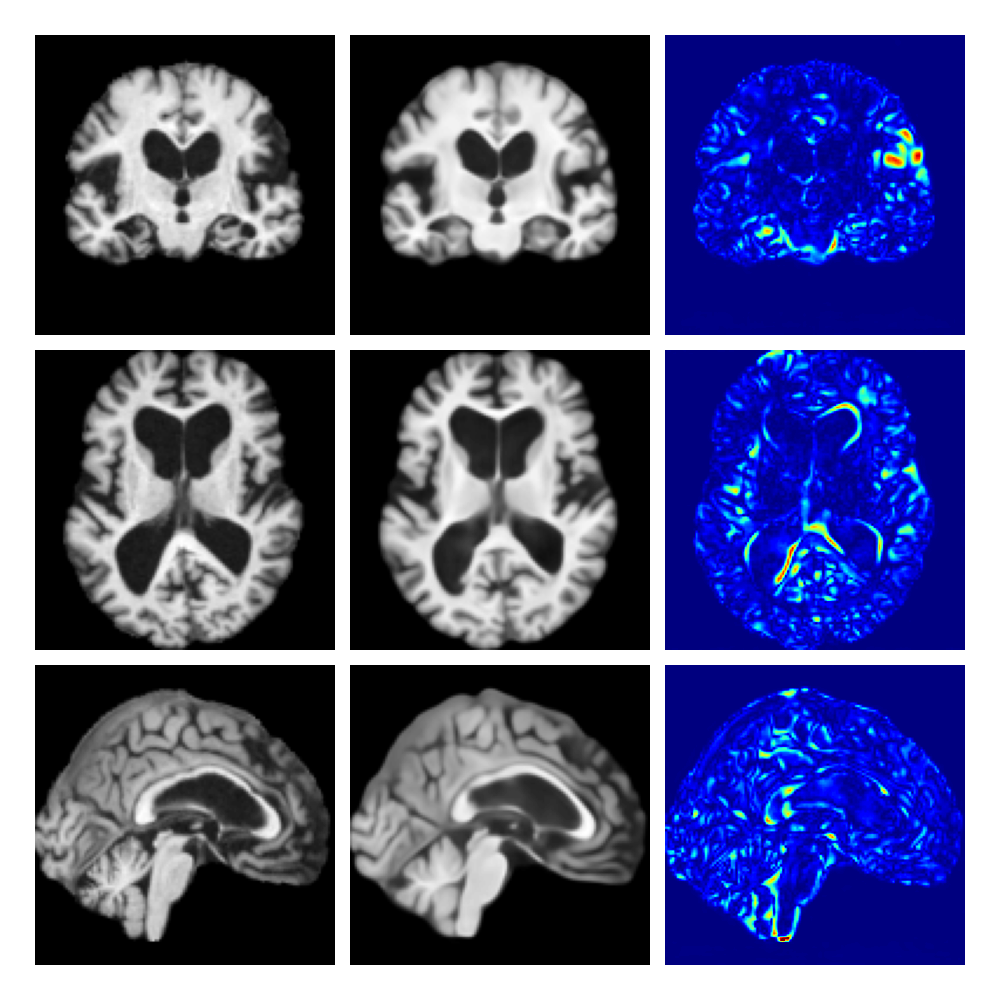}
        \caption{CADD (Ours)}
    \end{subfigure}
    
    \caption{Example reconstructions and anomaly maps for an AD sample from the disease cohort of the ADNI dataset. We expect to see some inpainting of atrophied tissue whilst retaining the defining characteristics of the individual sample.}
    \label{fig:example_ADNI_results}
\end{figure*}

Figures \ref{fig:enlarged_example_UKBB_results} and \ref{fig:enlarged_example_UoT_results} are enlarged versions of Figures \ref{fig:example_UKBB_results} and \ref{fig:example_UoT_results} respectively, with the latter now including results from all compared methods. Figure \ref{fig:example_ADNI_results} provides example reconstructions and anomaly maps for an AD subject from the ADNI disease cohort.

\begin{figure*}
    \centering
    \begin{subfigure}{0.12\linewidth} 
        \centering
        \includegraphics[width=\linewidth, trim=5 5 666 30, clip]{figures/ukbb/disease/5938271_CADD_thresh06.png}
        \caption{Input}
    \end{subfigure}
    \begin{subfigure}{0.24\linewidth} 
        \centering
        \includegraphics[width=\linewidth, trim=245 5 5 30, clip]{figures/ukbb/disease/5938271_VAE_thresh06.png}
        \caption{VAE}
    \end{subfigure}
    \begin{subfigure}{0.24\linewidth}
        \centering
        \includegraphics[width=\linewidth, trim=245 5 5 30, clip]{figures/ukbb/disease/5938271_cVAE_thresh06.png}
        \caption{cVAE}
    \end{subfigure}
    \begin{subfigure}{0.24\linewidth}
        \centering
        \includegraphics[width=\linewidth, trim=245 5 5 30, clip]{figures/ukbb/disease/5938271_LDMavg_thresh06.png}
        \caption{LDM ($T_{\text{avg}}$)}
    \end{subfigure}
    \begin{subfigure}{0.24\linewidth}
        \centering
        \includegraphics[width=\linewidth, trim=245 5 5 30, clip]{figures/ukbb/disease/5938271_LDM_thresh06.png}
        \caption{LDM}
    \end{subfigure}
    \begin{subfigure}{0.24\linewidth}
        \centering
        \includegraphics[width=\linewidth, trim=245 5 5 30, clip]{figures/ukbb/disease/5938271_AutoDDPM_thresh06.png}
        \caption{AutoDDPM}
    \end{subfigure}
    \begin{subfigure}{0.24\linewidth}
        \centering
        \includegraphics[width=\linewidth, trim=245 5 5 30, clip]{figures/ukbb/disease/5938271_THOR_thresh06.png}
        \caption{THOR}
    \end{subfigure}
    \begin{subfigure}{0.24\linewidth}
        \centering
        \includegraphics[width=\linewidth, trim=340 5 5 30, clip]{figures/ukbb/disease/5938271_CADD_thresh06.png}
        \caption{CADD (Ours)}
    \end{subfigure}
    
    \caption{Enlarged example reconstructions and anomaly maps for a sample from the disease cohort of the UKBB dataset. Lesion and WMH are indicated in the original image by the red and yellow boxes respectively.}
    \label{fig:enlarged_example_UKBB_results}
\end{figure*}

\end{document}